\documentclass[twocolumn]{aastex701}

\usepackage{amsmath}
\usepackage{mathtools}
\usepackage{xspace}
\usepackage{hyperref}

\newcommand{\figref}[1]{Figure~\ref{#1}}

\newcommand{\rg}{r_{\rm g}}
\newcommand{\tg}{t_{\rm g}}

\shorttitle{Slow-Light Effect in the Jet-Launching Region of M87}
\shortauthors{Y.~Tsunetoe et al.}

\begin{document}

\title{Slow-Light Effect in the Jet-Launching Region of M87}

\correspondingauthor{Yuh Tsunetoe}
\email{ytsunetoe@shao.ac.cn}

\author[0000-0003-0213-7628]{Yuh Tsunetoe}
\affiliation{Shanghai Astronomical Observatory, Chinese Academy of Sciences, Shanghai 200030,  P. R. China}
\affiliation{Center for Computational Sciences, University of Tsukuba, \ 1-1-1 Tennodai, Tsukuba, \ Ibaraki 305-8577, Japan}
\email{}

\author[0000-0002-5278-9221]{Dominic~W.~Pesce}
\affiliation{Center for Astrophysics $|$ Harvard \& Smithsonian, 60 Garden Street, Cambridge, MA 02138, USA}
\affiliation{Black Hole Initiative at Harvard University, 20 Garden Street, Cambridge, MA 02138, USA}
\email{}

\author[0000-0002-1919-2730]{Ramesh~Narayan}
\affiliation{Center for Astrophysics $|$ Harvard \& Smithsonian, 60 Garden Street, Cambridge, MA 02138, USA}
\affiliation{Black Hole Initiative at Harvard University, 20 Garden Street, Cambridge, MA 02138, USA}
\email{}



\begin{abstract}

We explore the impact of ``slow-light'' radiative transfer -- i.e., general relativistic radiative transfer (GRRT) calculations in which the simulated fluid evolves while light rays are propagating through it -- in general relativistic magnetohydrodynamic (GRMHD) models of the M87 jet.  Because the plasma in the jet-launching region is accelerated to relativistic velocities, and because the jet in M87 is nearly aligned with the line of sight (offset by $\sim$17\,degrees), a slow-light treatment is important for accurately modeling the observable structure.  While fast-light images exhibit prominent helical or loop-shaped features in the jet -- which we associate with narrow bundles of magnetic field lines -- these features become stretched and smoothed-out in slow-light images.  Our slow-light images instead exhibit a double-edged, cone-like morphology that is more consistent with observations of M87 than corresponding fast-light images.  We find that the radius at which the plasma transitions from sub-relativistic to relativistic velocities is imprinted on slow-light images via a transition from loop-dominated at small distances from the black hole to edge-dominated at a larger distance, with the loop–edge transition occurring at larger distances for lower black hole spins.  The jet image dynamics also vary with black hole spin, with low-spin models producing jets that exhibit substantial ``wobbling'', while high-spin models produce jets that are straighter and more stable in time.  The spin-dependent jet morphology and variability are revealed by slow-light imaging because slow-light effects become more enhanced as the plasma velocity becomes more relativistic, and because the plasma acceleration is itself a strong function of the spin.

\end{abstract}


\section{Introduction} \label{sec:intro}

{Astrophysical plasmas with magnetization parameter $\sigma \equiv B^2/4\pi \rho c^2 > 1$,  e.g., in the vicinity of neutron stars and black holes (BHs), and in relativistic jets, can experience strong accelerations.} 
These plasmas often end up with bulk speeds comparable to the speed of light, {leading to unusual relativistic effects} in the radiation observed on Earth, as 
exemplified for instance by the superluminal motion of jet blobs, in which the plasma appears to move faster than the speed of light {in the plane of the sky} 
\citep{1966Natur.211..468R,1981Natur.290..365P,1995ApJ...447..582B}.

The effects of relativistic plasma bulk motion have attracted considerable attention in the context of direct imaging of supermassive black holes (SMBHs), which has recently been realized with global-scale very long baseline interferometry (VLBI) at millimeter wavelengths \citep{2019ApJ...875L...1E,2022ApJ...930L..12E,2023Natur.616..686L}. 
In modeling radiation from the region near the BH, general relativistic radiative transfer (GRRT) calculations {\citep[e.g.,][]{2006MNRAS.367..905B,2009ApJ...697.1164B,2010ApJ...717.1092D,2016MNRAS.462..115D,Moscibrodzka2018,2020PASJ...72...32T,TsuY:2023,Prather2023}} are employed to produce theoretically predicted BH images based on plasma fluid models such as those obtained from general relativistic magnetohydrodynamic (GRMHD) simulations {\citep[e.g.,][]{Gammie2003,2019ApJ...875L...5E}}.

To account for the effects of relativistic motion of the emitting plasma (in other words, the finiteness of the speed of light), GRRT image calculations must be performed with the plasma fluid evolving concurrently with the propagation of light rays, a method known as the ``slow-light'' approach. 
Previous studies have compared slow-light images with those obtained using the ``fast-light'' approximation, which assumes that the plasma is frozen during light propagation (or equivalently, that the speed of light is infinite), and have found that the fast-light approximation is adequate for imaging the plasma in the  innermost regions around the BH, where the plasma bulk motion remains subrelativistic (e.g., \citealp{2010ApJ...717.1092D,2018A&A...613A...2B,2021MNRAS.508.4282M,2022ApJS..262...28W}; see also \citealp{2024A&A...689A.112V,2025arXiv251118410T} for modeling time lag in light curve). 
This finding has motivated many researchers to adopt the fast-light approximation, which allows the calculation of an image from a single GRMHD snapshot, making it computationally much less expensive than the slow-light approach, which requires combining a large number of GRMHD snapshots to generate a single slow-light image.

The slow-light treatment is, however, considered important for studying the jet-launching region, where the plasma is strongly accelerated and becomes relativistic. 
M87, one of the primary targets of high-resolution VLBI observations for near-BH imaging, exhibits a large-scale jet showing superluminal motion \citep[e.g.,][]{1995ApJ...447..582B,1999ApJ...520..621B,2007ApJ...663L..65C,2014ApJ...781L...2A,2018ApJ...855..128W,2019ApJ...887..147P} and hosts the SMBH M87*, which shows substantial variability in observed images on both day- and year-long timescales \citep{2020ApJ...901...67W,2022NatAs...6..259A,2022ApJ...930L..21B,2025arXiv250924593T}. 
Event Horizon Telescope (EHT) observations support magnetically arrested disk (MAD) GRMHD models for M87* \citep{2019ApJ...875L...5E,2021ApJ...910L..13E}, which are characterized by magnetic fields strong enough to influence the dynamics of accreting and ejecting plasmas around the BH \citep{Bisnovatyi1974, Igumenshchev2003, 2003PASJ...55L..69N, Tchekhovskoy2011}. 
MAD and other GRMHD models with a spinning BH predict acceleration of the plasma bulk speed in the jet from subrelativistic to relativistic regimes over scales of $\sim 10$~-~$1000~r_{\rm g}$ \citep{2006MNRAS.368.1561M, Tchekhovskoy2011, 2013MNRAS.436.3741P,2018ApJ...868..146N}, corresponding to sub-milliarcsecond (mas) scales in M87.\footnote{For M87, $1000~r_{\rm g}$ corresponds to $\approx 290~r_{\rm g}$ in projection on the sky plane for an inclination angle of $i = 163^\circ$ ($17^\circ$) with respect to the jet axis, which is $\approx 1.1~{\rm mas}$, assuming a BH mass $M_\bullet = 6.2\times10^9\,M_\odot$. Here $M_\odot$ is the Solar mass.} 
Here, $r_{\rm g} = GM_\bullet/c^2$ is the gravitational radius for a BH mass $M_\bullet$, where $G$ and $c$ are the gravitational constant and the speed of light, respectively. 
This makes M87 an excellent laboratory for testing the Blandford-Znajek process \citep{1977MNRAS.179..433B} as the central mechanism powering the BH jet and for constraining the BH spin parameter $a_*$.

In this work, we aim to demonstrate that the slow-light effect plays a significant role in the jet-launching region and to examine how it influences the {predicted} image features of the jet in M87. 
In our previous work \citep{2025ApJ...984...35T}, we showed that anisotropic nonthermal electrons can produce limb-brightened jet images from the jet-launching region to galactic scales, in good agreement with observations of M87. 
Building on this anisotropy model, we further investigate the imprints of magnetic field geometry and plasma acceleration profiles on the resulting images, with particular focus on their dependence on the BH spin. 
The resulting images provide theoretical predictions that can be directly compared with and tested by future VLBI observations at higher angular resolution and sensitivity, e.g., by the EHT \citep{2024arXiv241002986T}, the next-generation EHT (ngEHT; \citealp{2023Galax..11..107D}), and the Black Hole Explorer (BHEX; \citealp{2024SPIE13092E..2DJ,2024SPIE13092E..2EA}).

\bigskip

The structure of this paper is as follows.
The method for calculating slow-light images via GRRT based on the MAD GRMHD model is described in Section~\ref{sec:method}.
In Section~\ref{sec:results}, we examine the effects of the slow-light treatment on image features by comparing the calculated slow-light images with conventional fast-light images.
We then extend our analysis to survey slow-light effects across different black hole spins and at multiple wavelengths.
Section~\ref{sec:discussion} discusses the relationship between the slow-light images, the magnetic field geometry in the jet, and the black hole spin, as well as the consistency of our calculated images with observations of M87.
Finally, Section~\ref{sec:conclusions} presents our conclusions.

\section{Method} \label{sec:method}

\subsection{MAD GRMHD Simulations and Anisotropic GRRT Modeling}\label{subsec:model}

\begin{table*}\label{table:Mdot_h}
\begin{center}
  \begin{tabular}{c||c|c|}
     $a_*$ & $\dot{M}$ & $h$ \\ \hline 
    0.9 & $5.2\times10^{-4}~M_\odot / {\rm yr}$ & 0.0025 \\ 
    0.7 & $1.1\times10^{-3}~M_\odot / {\rm yr}$ & 0.0025 \\ 
    0.5 & $1.3\times10^{-3}~M_\odot / {\rm yr}$ & 0.005 \\ 
    0.3 & $1.3\times10^{-3}~M_\odot / {\rm yr}$ & 0.05 \\
  \end{tabular}
\end{center}
  \caption{List of the mass accretion rate onto the BH ($\dot{M}$) and the energy injection efficiency into the nonthermal electrons ($h$) for the four BH spin models. 
  }
\end{table*}

We adopt the plasma fluid and radiation models developed in our previous work \citep{2025ApJ...984...35T}, which combine the MAD GRMHD simulation data of \citet{2022MNRAS.511.3795N} with a synchrotron emission model that incorporates an anisotropic, nonthermal, power-law electron distribution.

First, we demarcate the GRMHD model into disk and jet regions based on the magnetization parameter: $\sigma < 1$ for the disk and $1 \leq \sigma < 300/\sqrt{r}$ for the jet {(the upper limit is to avoid regions of the GRMHD solution that are unreliable because of numerical problems)}. 
We assume that the electron distribution is purely thermal in the disk, and purely nonthermal, with an anisotropic power-law form, in the jet region. 
For the thermal electrons in the disk, the two parameters in the $R$-$\beta$ prescription \citep{2016A&A...586A..38M} are set to $(R_{\rm low}, R_{\rm high}) = (10, 160)$ to determine the electron temperature from the gas temperature and plasma-$\beta$ value given by the GRMHD simulation.

In the jet region, we assume that the energy of nonthermal electrons injected into the jet plasma is proportional to the Poynting flux, 
\begin{equation}\label{eq:u_inj}
    u_{\rm nt,inj} = h\,\frac{|\mathbf{S}|}{c},
\end{equation}
where $\mathbf{S}$ is the Poynting flux in the GRMHD simulation\footnote{{The Poynting flux is computed in the zero angular momentum observer (ZAMO) frame.}} and $h$ is the energy injection efficiency. 
We take the anisotropic power-law distribution function to be given by
\begin{equation}\label{eq:aniso}
    f(\gamma,\xi) = \phi(\xi)\, f_{\rm iso}(\gamma),
\end{equation}
\begin{eqnarray}
   f_{\rm iso}(\gamma) =
   \begin{dcases}
      0, & \gamma < \gamma_{\rm min}, \\
      \dfrac{n_{\rm pl}(p-1)}{\gamma_{\rm min}^{1-p}-\gamma_{\rm max}^{1-p}}\,\gamma^{-p}, & \gamma_{\rm min} \leq \gamma \leq \gamma_{\rm max}, \\ 
      0, & \gamma > \gamma_{\rm max},
   \end{dcases}
\end{eqnarray}
\begin{equation}\label{eq:phi}
    \phi(\xi) = P(p,\eta)^{-1}\left[1 + (\eta - 1)\cos^2\xi \right]^{-p/2},
\end{equation}
\begin{equation}
    P(p,\eta) = \frac{1}{2}\int_0^{\pi} \!{\rm d}\xi\, \sin\xi\,\left[1 + (\eta - 1)\cos^2\xi \right]^{-p/2}.
\end{equation}
Here, $\gamma$ and $\xi$ denote the Lorentz factor and pitch angle, $p$ is the power-law index, $\gamma_{\rm min}$ and $\gamma_{\rm max}$ are the minimum and maximum Lorentz factors, and $\eta$ is a measure of the pitch-angle anisotropy. $n_{\rm pl}$ represents the number density of power-law electrons. 
{The injected electron population is then permitted to cool via synchrotron radiation, resulting in a broken double power-law distribution with two indices $p_1$ and $p_2$; see the Appendix of \citealt{2025ApJ...984...35T} for details.}

We focus on modeling the M87 jet and set the BH mass to $M_\bullet = 6.2\times10^9~M_\odot$ \citep{2019ApJ...875L...1E} and the observer's inclination angle to $i = 163^\circ$ \citep{2018ApJ...855..128W} with respect to the jet axis (equivalently, the $z$-axis in the GRMHD simulation coordinates). 
We determine the two power-law indices, $p_1$ and $p_2$, and the minimum and break Lorentz factors, $\gamma_{\rm min}$ and $\gamma_{\rm br}$, in the broken power-law distribution based on the synchrotron cooling prescription (see the Appendix of \citealt{2025ApJ...984...35T}), while fixing the maximum Lorentz factor to $\gamma_{\rm max} = 10^8$ and the anisotropy parameter to $\eta = 0.01$ in the jet. 
The GRRT calculations are performed using the \texttt{SHAKO} code (\citealt{TsuY:2023}; see also the Appendix of \citealt{2024PASJ...76.1211T} for validation of the code). 

In this work, we use GRMHD data from \cite{2022MNRAS.511.3795N} corresponding to four BH spin values, $a_* = 0.3,\, 0.5,\, 0.7,\, \text{and}\, 0.9$. 
For each spin, we scale the mass accretion rate $\dot{M}$ in the GRMHD model (this is allowed since GRMHD models are scale-invariant) such that the thermal synchrotron-emitting electrons in the disk produce an average total flux of $\sim 0.5~{\rm Jy}$ at 230~GHz. 
After fixing $\dot{M}$, we set the nonthermal energy injection efficiency factor $h$ for the jet plasma such that the combined emission from the thermal electrons in the disk and the nonthermal electrons in the jet gives an average total flux of $\sim 1~{\rm Jy}$ at 86~GHz. 
The sets of obtained $\dot{M}$ and $h$ are listed in Table \ref{table:Mdot_h}. 
The two models with $a_*=0.5$ and $0.9$ show consistent values of $\dot{M}$ with the image modeling studies of M87* performed in \cite{2021ApJ...910L..13E}, following the results in \cite{2025ApJ...984...35T}. 

\begin{figure*}
\begin{center}
	\includegraphics[width=19cm]{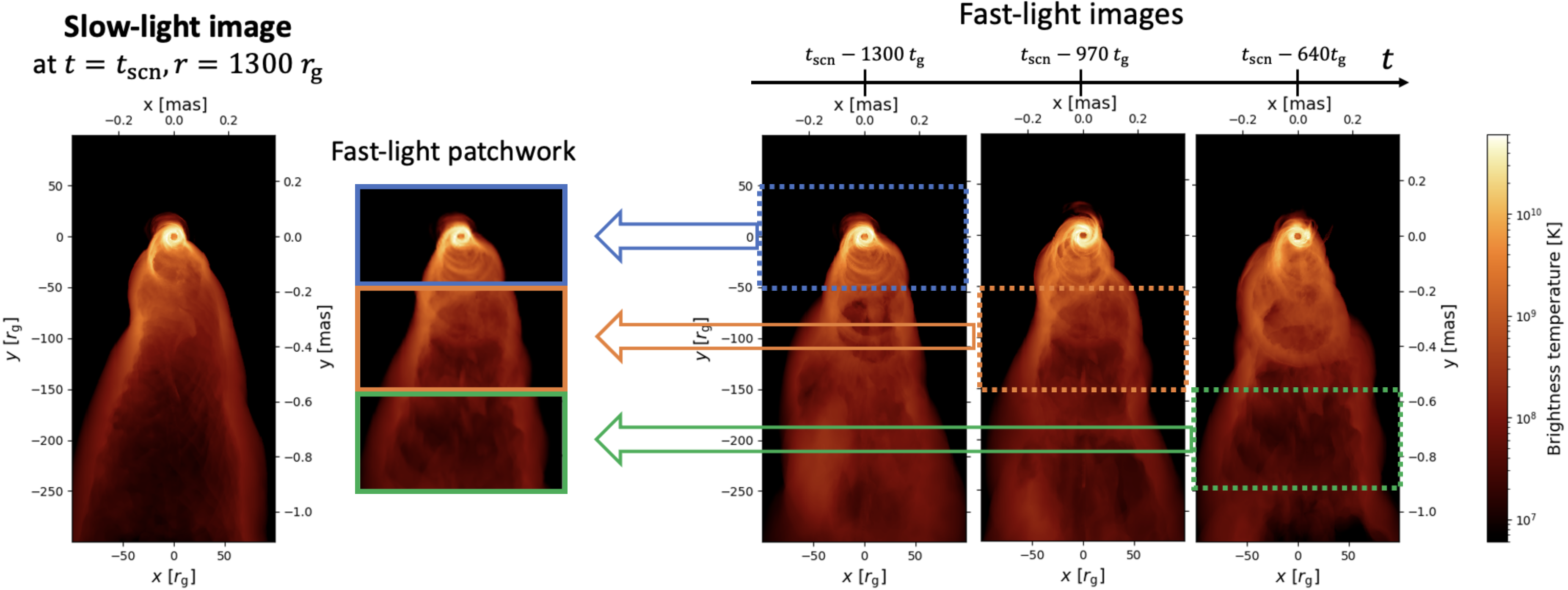}
\end{center}
    \caption{
    Left: an image at 86~GHz at $t = t_{\rm scn}$ ($= 22{,}250~t_{\rm g}$) obtained from the slow-light calculation for the high-spin case ($a_* = 0.9$). 
    Right: three images from the fast-light calculations based on the GRMHD snapshots at $t = t_{\rm scn} - 1300~t_{\rm g}$, $t_{\rm scn} - 970~t_{\rm g}$, and $t_{\rm scn} - 640~t_{\rm g}$ (left to right). 
    These three epochs correspond to the times when the light rays reaching $y = 0$, $-100~r_{\rm g}$, and $-200~r_{\rm g}$ in the slow-light image pass through the jet. 
    The corresponding regions are extracted from the fast-light images and combined in a patchwork form (second panel from left) aligned with the slow-light image.
        }
    \label{fig:slow_fast}
\end{figure*}

\subsection{Slow-Light GRRT Calculations}\label{subsec:slowlight}

In the slow-light approach, the plasma in the GRMHD simulation evolves while light rays propagate toward the observer's screen in the GRRT calculation, as mentioned in the Introduction. 
To implement this, we place the observer’s screen, with a size of $200~r_{\rm g} \times 400~r_{\rm g}$ and $224\times448$ pixels, at $r = 1300~r_{\rm g}$, and compute the trajectories of light rays for each pixel backward in time, using general relativistic ray tracing. 
We set the Kerr-Schild time coordinate $t = t_{\rm scn}$ at the screen for each ray and trace it back to $t = t_{\rm scn} - 2000~t_{\rm g}$, beyond which we confirmed that the contribution to the observed intensity becomes negligible in our setup. 
Here, $t_{\rm g} = r_{\rm g}/c = GM_\bullet/c^3$ is the light-crossing time of one gravitational radius. 
Once the light paths are determined, radiative transfer is computed {in the forward direction, from $t_{\rm scn} - 2000~t_{\rm g}$ to $t_{\rm scn}$,} using at each step in the integration the closest GRMHD snapshot data in time. 
The GRMHD datasets are sampled at a cadence of $2~t_{\rm g}$ (see Appendix~\ref{apdx:cad} for a discussion of the cadence).

We calculate image movies over a duration of $5000~t_{\rm g}$ for $t_{\rm scn} = [22{,}000~t_{\rm g}, 27{,}000~t_{\rm g}]$ to determine $\dot{M}$ and $h$ from the average total fluxes at 86 and 230~GHz, as described in the final paragraph of Subsection~\ref{subsec:model}. 
In addition to the slow-light image and movie calculations which are the main goal of this work, we also perform corresponding image and movie calculations under the fast-light approximation, which are used in Section~\ref{sec:results} to compare and discuss the image features produced by the slow-light treatment.

\section{Results} \label{sec:results}

In this section, we first present the effects of the slow-light treatment on the images for our highest-spin model, $a_* = 0.9$, which exhibits the most pronounced differences from fast-light because it has the strongest acceleration. 
We then examine the slow-light effects across the four spin models, focusing on the differences in their plasma acceleration profiles.

\subsection{Slow-Light Image and Fast-Light Patchwork}\label{subsec:slow_fast}

\begin{figure*}
\begin{center}
	\includegraphics[width=10cm]{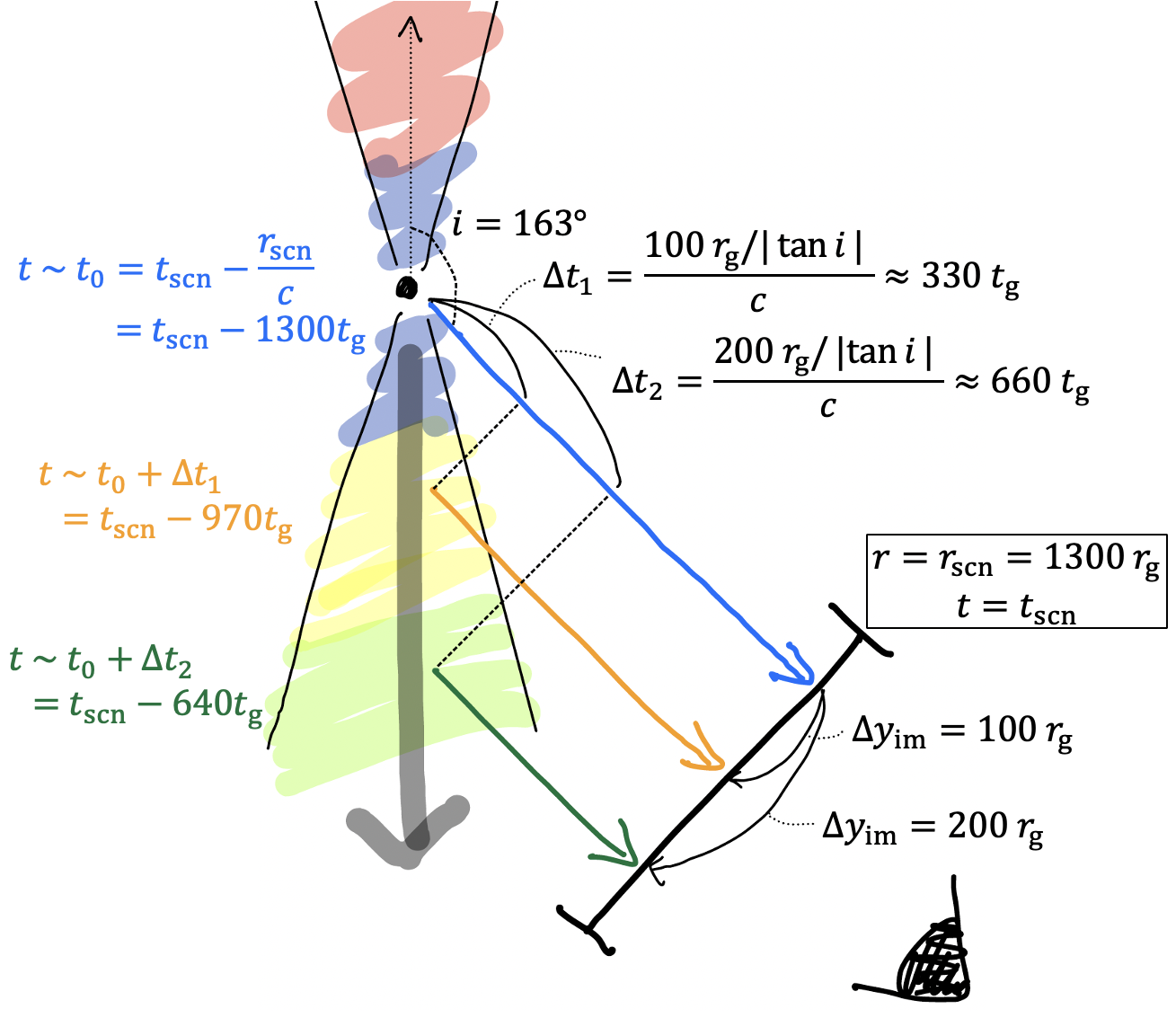}
\end{center}
    \caption{
    A schematic illustration showing the correspondence between a slow-light image and a set of fast-light images, as shown in Figure \ref{fig:slow_fast}.  
    The blue, orange, and green arrows represent the patchwork shown in Figure~\ref{fig:slow_fast}. 
    The blue ray passes through the jet or around the BH at $t = t_0 \sim t_{\rm scn} - 1300~t_{\rm g}$ and reaches the origin of the observer's screen at $r_{\rm scn} = 1300~r_{\rm g}$ and $t = t_{\rm scn}$. 
    The orange (green) ray, which reaches $y = -100~r_{\rm g}$ ($-200~r_{\rm g}$) on the screen, passes near the jet axis at approximately $t \sim t_0 + \Delta t_1$ ($t_0 + \Delta t_2$). 
    Here, the time delay $\Delta t_1$ ($\Delta t_2$) is geometrically obtained by tangentially deprojecting the corresponding distance $100~\rg$ ($200~\rg$) on the screen onto the light ray for an inclination of $i = 163^\circ$. 
    This indicates that the light rays in the downstream region of the jet originate from later epochs. 
    }
    \label{fig:pictures}
\end{figure*}

A snapshot image at 86~GHz from the slow-light calculation is shown on the left of \figref{fig:slow_fast}, accompanied by three images obtained with the fast-light approximation on the right. 
In principle, there is no one-to-one correspondence between any pair of snapshots from the slow- and fast-light calculations. 
As described in Subsection~\ref{subsec:slowlight}, the slow-light images are computed with the GRMHD model evolving over the duration $[t_{\rm scn} - 2000~t_{\rm g},\, t_{\rm scn}]$ spanned by the ray integration, whereas each fast-light image is based on a single GRMHD snapshot at a fixed time. 
This explains why the overall structure of the slow-light image on the left does not coincide with that of any fast-light image on the right.

\begin{figure*}
\begin{center}
    \includegraphics[width=10cm]{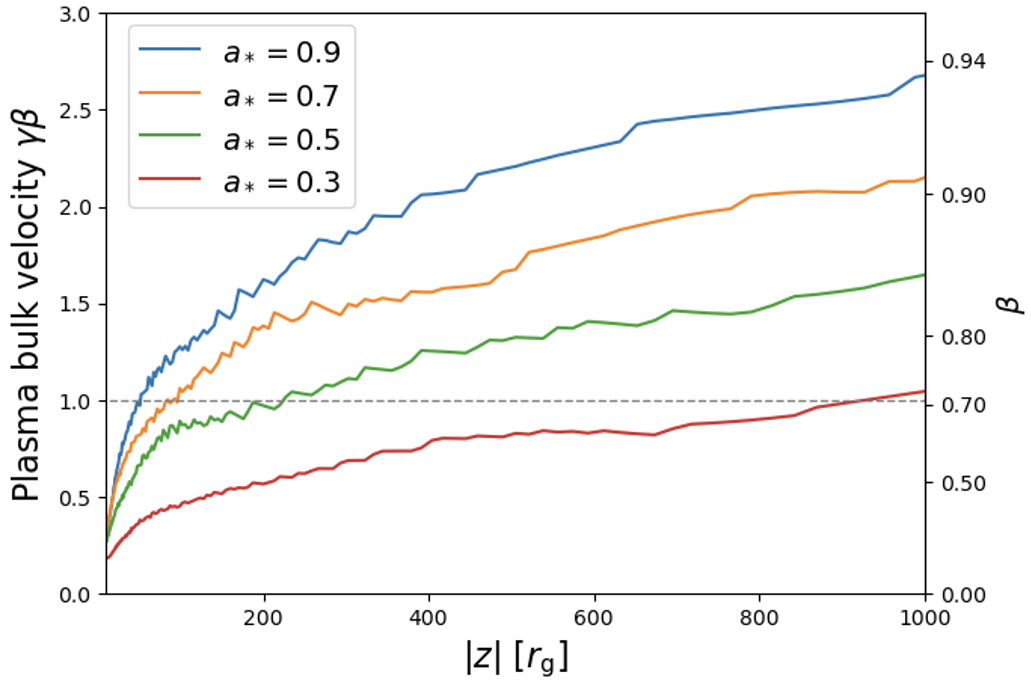}
\end{center}
    \caption{
    Profiles of the modified plasma bulk velocity, $\gamma\beta$, along the $z$-axis for the four spin models. 
    Here, $\beta = v/c$ is the plasma bulk velocity normalized by the speed of light, and $\gamma = 1/\sqrt{1 - \beta^2}$ is the corresponding Lorentz factor. 
    Note that $\gamma\beta = \sqrt{\gamma^2 - 1}$ ranges from 0 to infinity. 
    The velocity is calculated within the jet region defined by $1 < \sigma < 300/\sqrt{r}$, using the GRMHD data averaged over $5000~t_{\rm g}$. 
    Tick marks on the right side of the panel indicate the corresponding values of $\beta$.
    }
    \label{fig:velo_prof}
\end{figure*}

Nevertheless, we can find fast-light images that resemble different regions of a single slow-light snapshot. 
For instance, the leftmost fast-light image in \figref{fig:slow_fast} (the third panel from right) is calculated based on the GRMHD snapshot at $t = t_{\rm scn} - 1300~t_{\rm g}$. 
It shows a similar structure in the inner region, $-50~r_{\rm g} < y < 50~r_{\rm g}$, to that seen in the corresponding part of the slow-light image on the left. 
This is because the light rays for the pixels in this region pass through or near the BH at around $t \sim t_{\rm scn} - 1300~t_{\rm g}$ before reaching the screen at $r = 1300~r_{\rm g}$, as shown by the blue arrow in Figure \ref{fig:pictures}. 
This result reproduces the findings of previous studies, which concluded that the fast-light approximation works well for calculating the innermost BH images. 
Meanwhile, we can also see that the slow-light image begins to deviate for $y \lesssim -30~r_{\rm g}$, since the plasma in the jet is accelerated and becomes relativistic at $|z| \sim 30~r_{\rm g} / {\sin}(163^\circ) \approx 100~r_{\rm g}$,\footnote{This estimate is based on the simple assumption that the synchrotron-emitting plasma propagates rectilinearly along the $z$-axis. 
In our actual models, however, nonthermal electrons are preferentially injected into the jet sheath region and the plasma follows a helical trajectory.} as shown in \figref{fig:velo_prof}.

Second, we also find a similarity in the downstream jet region of $-150~r_{\rm g} < y < -50~r_{\rm g}$ between the slow-light image at $t = t_{\rm scn}$ and the middle panel of the fast-light image set, which is based on the GRMHD snapshot at $t = t_{\rm scn} - 970~t_{\rm g}$. 
For example, a prominent waist-like feature in the jet can be seen around $y \sim -80~r_{\rm g}$, which is also visible in the fast-light image.
This can be naturally understood from the fact that the light ray reaching $(x,y) = (0,-100~r_{\rm g})$ on the image passes near the $z$-axis about $100~r_{\rm g} / c / |{\rm tan}(163^\circ)| \approx 330~t_{\rm g}$ later than the light ray reaching $(x,y) = (0,0)$, as described by the orange arrow in \figref{fig:pictures}. 
During this time, the jet propagates outward by $\sim 330~t_{\rm g} \times \beta \approx 300~r_{\rm g}$, assuming a bulk velocity of $\beta \sim 0.9$ as shown in \figref{fig:velo_prof}. 
This propagation distance corresponds to $\sim 300~r_{\rm g} \sin(163^\circ) \approx 90~r_{\rm g}$ in the downstream direction on the image.

Finally, a similar correspondence can be found between the outer jet region, $-250~r_{\rm g} < y < -150~r_{\rm g}$, in the slow-light image and the rightmost panel of the fast-light image set at $t = t_{\rm scn} - 640~t_{\rm g}$, as denoted by the green arrow in \figref{fig:pictures}. 
As in the above case, the light rays passing through the outer jet do so about $660~t_{\rm g}$ later than those passing near the BH, during which the jet propagates about $180~r_{\rm g}$ downward on the image. 
In the fast-light image, the jet exhibits a transverse expansion and nearly reaches the edges of the screen — a feature that is also visible in the slow-light image.

Building on the discussion above, we can construct a patchwork image from the three fast-light images, 
shown as the second panel from the left in \figref{fig:slow_fast}, where the more downstream portions of the patchwork are taken from later snapshot images. 
The patched image shows good agreement with the slow-light image in the overall shape of the jet, despite the rough procedure used to reconcile the slow-light and fast-light methods.\footnote{This shows that, in an observed image of the M87 jet on the sub-mas scale, the radiation we see in the downstream region was generated later by about $660~\tg \sim $ eight months, compared to the ring emission around the BH.} 
This also implies that the relativistic jet appears to propagate more rapidly in the slow-light movie than in the fast-light case, manifesting the superluminal motion effect. 
However, the agreement is less good in the interior regions of the jet: the slow-light image exhibits a clear two-edged morphology with a smooth interior, whereas all the fast-light images contain loop-shaped emission features. 
This difference is investigated in the next subsection.

\subsection{Two Edges {in Slow-Light vs} Loops in {Fast-Light}}\label{subsec:edges_loop}

\begin{figure*}
\begin{center}
	\includegraphics[width=19cm]{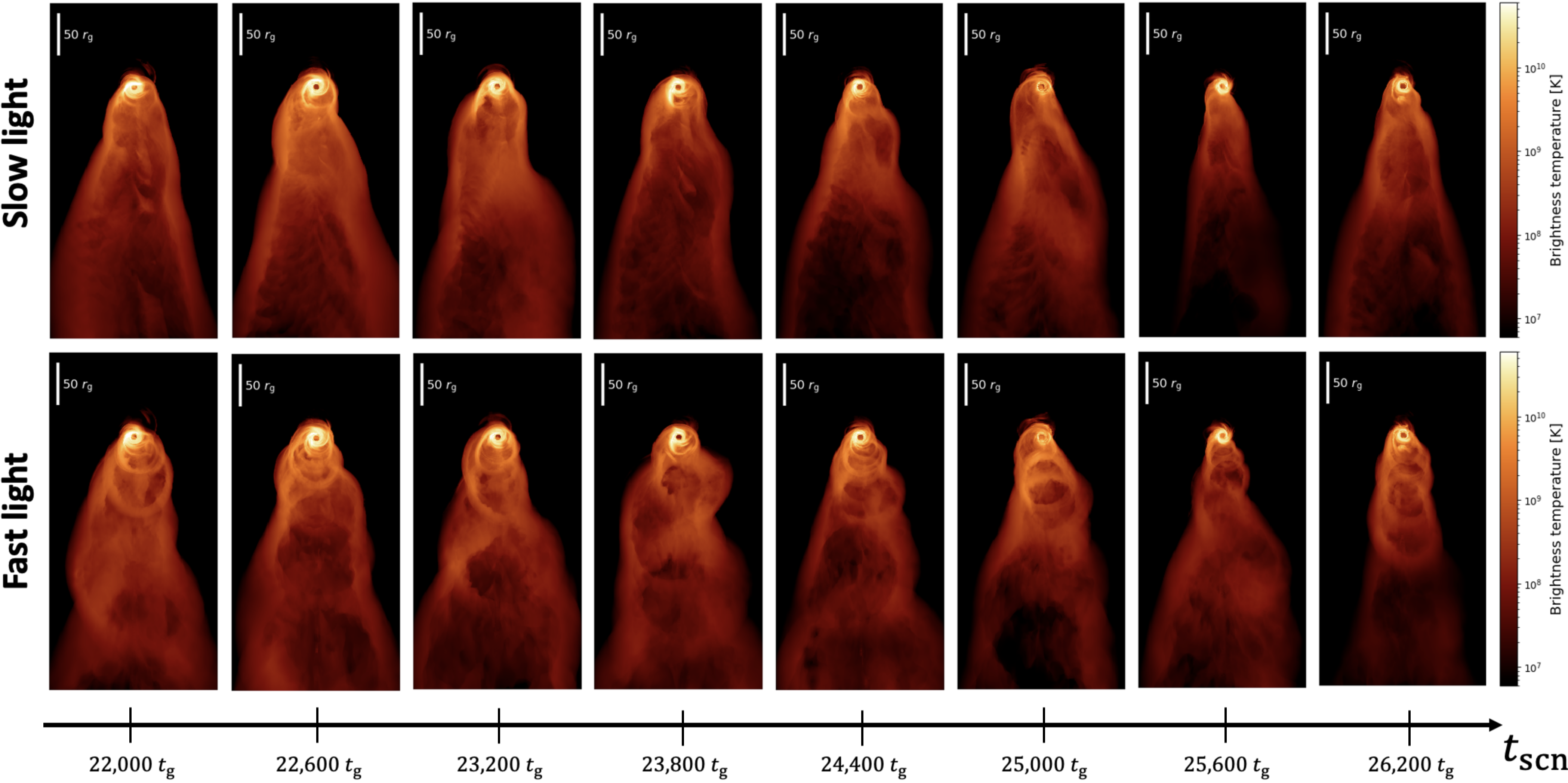}
\end{center}
    \caption{
    Eight snapshot images at 86 GHz from the slow-light (upper row) and fast-light (lower row) calculations for the GRMHD model with $a_* = 0.9$. 
    The slow-light images are taken from $t_{\rm scn} = 22,000~t_{\rm g}$ to $26{,}200~t_{\rm g}$ with a cadence of $600~t_{\rm g}$. 
    The fast-light images are based on the GRMHD snapshot data at $t = t_{\rm scn} - 1300~t_{\rm g}$ (see also Figure~\ref{fig:slow_fast}). 
    Movies are available in the online article and on \href{https://youtu.be/VIzLOhfJFaI}{YouTube}.
        }
    \label{fig:8snapshots}
\end{figure*}

In \figref{fig:8snapshots}, eight slow-light images from $t_{\rm scn} = 22{,}000~t_{\rm g}$ to $26{,}200~t_{\rm g}$ are shown at a cadence of $600~t_{\rm g}$, together with fast-light images generated from the corresponding GRMHD snapshots at $t = t_{\rm scn} - 1300~t_{\rm g}$. 
Each pair of slow- and fast-light images exhibits similar features in the inner region around the BH, as discussed in the previous subsection. 
At the same time, the slow-light images tend to display a persistently limb-brightened (two-edge) morphology, whereas the fast-light images show more helical or loop-like structures.

The sharper jet edges seen in the slow-light images can be interpreted as a consequence of the nearly parallel motion between the jet plasma flow and the emitted light rays {($17^\circ$ between the two directions in M87 assuming vertical flow)}. 
As discussed in the previous subsection and illustrated in \figref{fig:pictures}, a slow-light image can be regarded approximately as a patchwork of fast-light images taken at different times.

In the slow-light treatment, jet propagation in a direction nearly parallel to that of light propagation also affects the radiative transfer process along a given light ray. 
A relativistic jet component that emits radiation at a certain spatial and temporal point can nearly keep up with the light ray and continuously contribute to the observed intensity along its trajectory, particularly for small inclination angles as in M87. 
These effects naturally lead to an apparent stretching and smoothing of the intrinsically helical jet components in GRMHD models, blurring the jet interior while at the same time reinforcing the jet edges in the slow-light images. 

\begin{figure*}
\begin{center}
	\includegraphics[width=13cm]{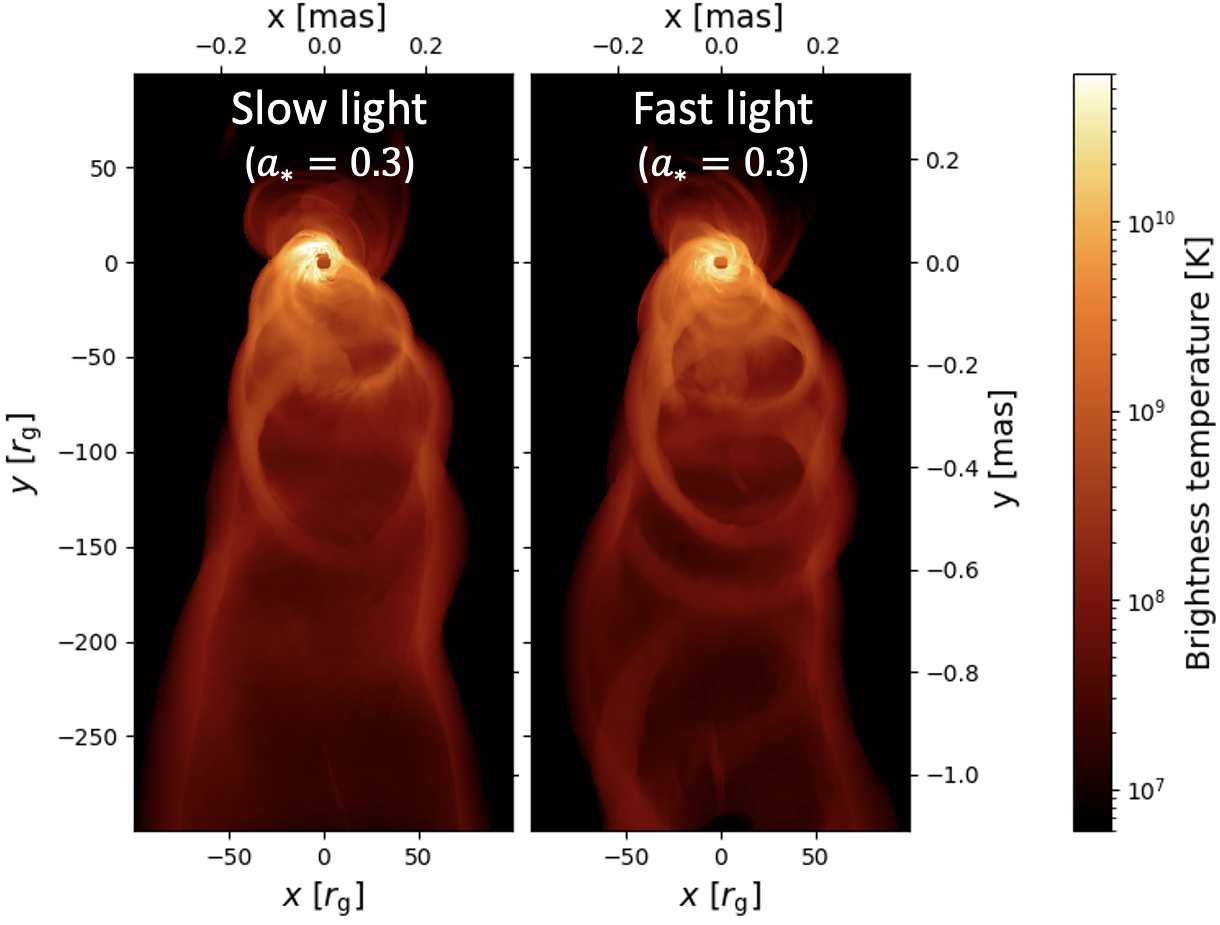}
\end{center}
    \caption{
    Snapshot images at 86 GHz by slow-light (left) and fast-light (right) calculations for the $a_* = 0.3$ model. The slow light image is taken at $t = t_{\rm scn} = 24,650~\tg$. The fast-light image is based on the GRMHD snapshot at $t = t_{\rm scn}-1015~\tg$. 
        }
    \label{fig:a03_slow_fast}
\end{figure*}

To further illustrate the effect using a model with a relatively mild slow-light effect, a set of slow- and fast-light images for the lowest-spin model with $a_* = 0.3$ is shown in Figure~\ref{fig:a03_slow_fast}. 
This model produces the weakest acceleration of the plasma, as shown in Figure~\ref{fig:velo_prof}. 
Here, the fast-light image, based on the GRMHD snapshot at $t \sim t_{\rm scn} - 1000~t_{\rm g}$, shows similar features to the slow-light image in the region around $y \sim -100~r_{\rm g}$ (see also Figure~\ref{fig:slow_fast}). 
In particular, both images exhibit a large loop-shaped emission with its upper edge located at $y \approx -40~r_{\rm g}$. 
However, the lower edge of the loop appears above $y = -150~r_{\rm g}$ in the fast-light image, whereas in the slow-light image the loop extends below $y = -150~r_{\rm g}$. 
Furthermore, the next loop, ranging from $y = -80~r_{\rm g}$ to $-170~r_{\rm g}$ in the fast-light image, is smoothed and stretched out to the region around $y = -120~r_{\rm g}$ to $-220~r_{\rm g}$ in the slow-light image. 
In particular, the lower part of this loop becomes blurred and almost invisible, forming a double-edged structure by merging with the upper half of the subsequent loop, which appears below $y = -170~r_{\rm g}$ in the fast-light image. 
In this way, we can clearly identify a mild but distinct slow-light effect that stretches and smooths the jet in the low-spin model, resulting in a transition from a loop-like to a double-edged morphology (see also Subsection~\ref{subsec:loop_edge}). 
In the highest-spin model ($a_* = 0.9$), this effect becomes significant already close to the BH due to the stronger plasma acceleration, producing a prominently double-edged structure nearer the jet base, as seen in Figure~\ref{fig:8snapshots}.

Similarly, \citet{2021A&A...653A..10M} performed slow-light image calculations based on small-scale particle-in-cell (PIC) simulations for an edge-on view ($i = 90^\circ$) and found that the slow-light treatment leads to an averaging of plasma emission features along each light path and to a blurring of the overall image morphology. 
The transformation from a helical to a conical appearance in our jet images can be interpreted as an extreme case of such averaging in the nearly face-on view (indeed, the fast-light image also shows a conical jet without any hint of loops after time-averaging; see the following subsection).\footnote{In this sense, the slow-light effect is analogous to motion blur in photography. While motion blur arises from the finite exposure time of the camera, the slow-light effect originates from the finite speed of light.}
Recently, \citet{2025A&A...693A.169S} also found a broadening of shocks in the jet of the large-inclination ($i = 80^\circ$) source NGC~1052, using slow-light images based on a special relativistic hydrodynamical (SRHD) model on several-mas scales. 
They also reported an asymmetry in a double-edge jet caused by the relativistic beaming (see also Appendix~\ref{apdx:iso} for asymmetric images with isotropic electrons).

Since the anisotropic emission is strongest along the two edges of the jet, where the helical magnetic field lines are well aligned with the direction of the light rays \citep{2025ApJ...984...35T}, the resulting conical jet images exhibit two bright edges that are more pronounced relative to the dimmer interior region than those seen in the looped jet structures of the fast-light images. 
Furthermore, transforming the loop-like structures into a hollow conical geometry naturally increases the line-of-sight optical depth along the cone’s edges, where the emitting path length is longer, making them brighter relative to the central ridge and enhancing the contrast across the jet.
In this way, the slow-light effect enhances the limb-brightening feature produced by anisotropic electrons in the relativistic jet, leading to a better agreement with the actual observations of M87 (e.g., \citealp{2018A&A...616A.188K,2018ApJ...855..128W,2023Natur.616..686L,2025A&A...696A.169K}). {Note that slow-light by itself does not produce limb-brightening; an anisotropic electron distribution is essential, as shown in} Appendix \ref{apdx:iso}.

\subsection{Convergence of Time-Averaged Images}\label{subsec:average}

\begin{figure*}
\begin{center}
	\includegraphics[width=13cm]{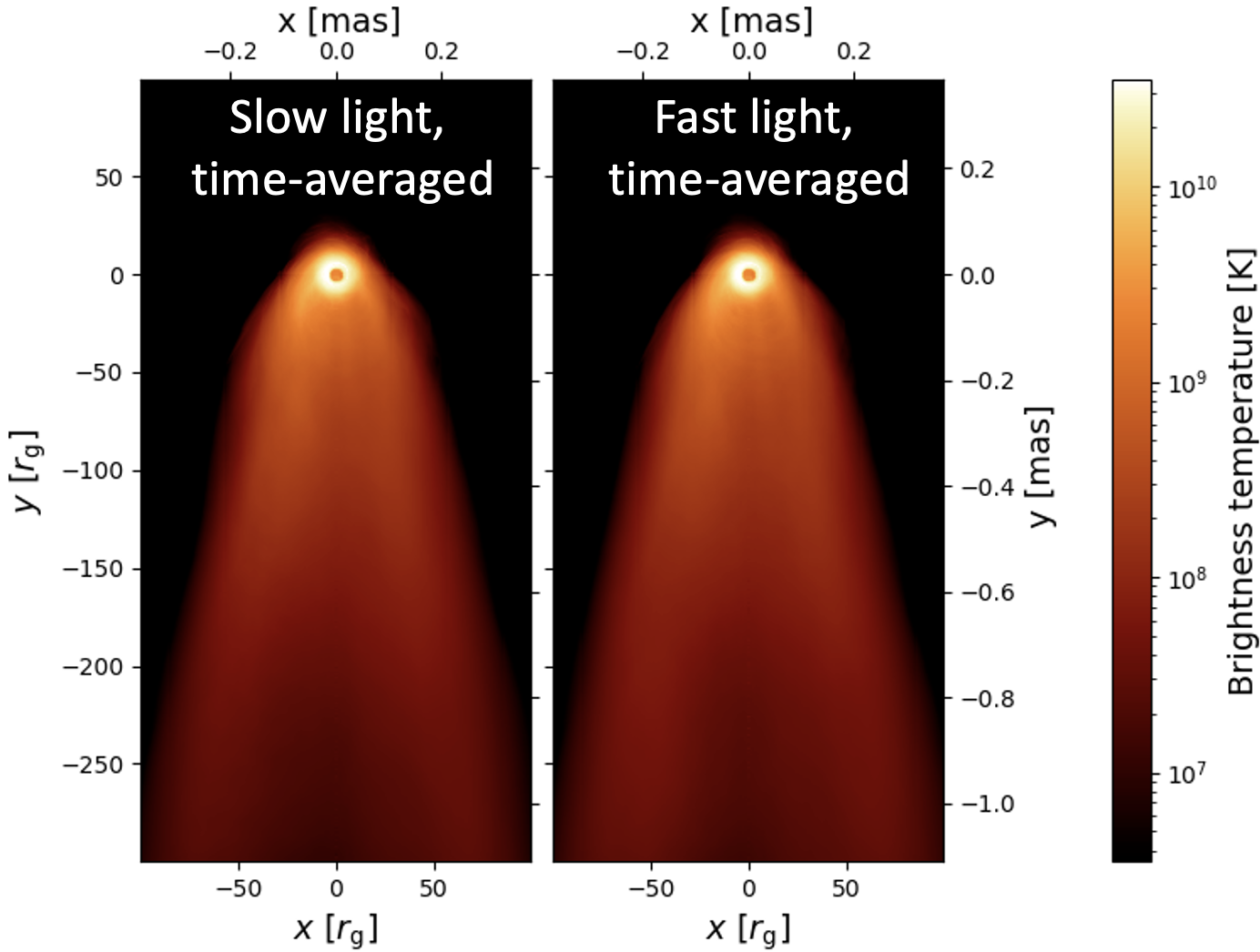}
\end{center}
    \caption{
    Time-averaged images at 86 GHz from the slow-light (left) and fast-light (right) calculations for the high-spin model with $a_* = 0.9$. 
    The slow-light image is obtained by averaging over a duration of $5000~t_{\rm g}$ for $t_{\rm scn} = [22,000~t_{\rm g}, 27,000~t_{\rm g}]$, with a cadence of $50~t_{\rm g}$. 
    The fast-light counterpart is averaged from snapshot images based on the GRMHD data for $t = [22,000~t_{\rm g} - 1300~t_{\rm g}, 27,000~t_{\rm g} - 1300~t_{\rm g}]$, using the same cadence.
        }
    \label{fig:average}
\end{figure*}

In the previous subsection, we showed that the slow-light treatment significantly affects snapshot images, producing sharper edge features than those found in fast-light images. 
Next, we examine the effect on time-averaged images.

In \figref{fig:average}, time-averaged images are shown for both the slow- and fast-light calculations. 
The slow-light image is obtained by averaging the movie frames over a duration of $5000~t_{\rm g}$ for $t_{\rm scn} = [22{,}000~t_{\rm g}, 27{,}000~t_{\rm g}]$, 
while the fast-light image is produced by averaging the images calculated from the corresponding GRMHD snapshots at $t = [22{,}000~t_{\rm g} - 1300~t_{\rm g}, 27{,}000~t_{\rm g} - 1300~t_{\rm g}]$. 
Despite the large differences seen in the snapshot images shown in \figref{fig:8snapshots}, the two time-averaged images are in excellent agreement, both showing a smooth, straight, double-edged jet. 
The average total fluxes at 86~GHz also agree very well between the two calculations—1.03~Jy for the slow-light case and 1.05~Jy for the fast-light case—serving as a sanity check of our slow-light implementation.

This good agreement in the time-averaged images arises from the fact that the inner region around the BH in both calculations is persistently dominated by generally optically-thin emissions generated at similar epochs, with an offset of $-1300~t_{\rm g}$ in the fast-light case, as shown in Figures~\ref{fig:slow_fast} and~\ref{fig:8snapshots}. 
A closer inspection of \figref{fig:average} reveals a small deviation in the outer jet region around $y \lesssim -200~r_{\rm g}$ between the two images. 
This difference occurs because the GRMHD evolution during the later epoch of $[27{,}000~t_{\rm g} - 1300~t_{\rm g}, 27{,}000~t_{\rm g}]$ is not included in the fast-light calculation but contributes to the outer jet emission at later times in the slow-light case.\footnote{We confirmed that a similar outer-jet structure can be reproduced in the fast-light calculation by introducing a small time offset, say, $[22{,}000~t_{\rm g} - 640~t_{\rm g}, 27{,}000~t_{\rm g} - 640~t_{\rm g}]$ (see also \figref{fig:slow_fast}). 
However, in that case, the inner image becomes inconsistent because the fast-light approach then misses the evolution at earlier epochs.}

\subsection{Loop–Edge Transition and Black Hole Spin}\label{subsec:loop_edge}

So far, we have seen that the slow-light effect plays a significant role in the images of the jet-launching region in M87, based on the highest-spin model with $a_* = 0.9$, which produces the strongest acceleration among our GRMHD models (\autoref{fig:velo_prof}) and is expected to exhibit the most pronounced difference from the fast-light case. 
Here, we extend our analysis to lower-spin models to examine the dependence of the slow-light features on the BH spin, focusing on the velocity profile of the plasma along the jet shown in \figref{fig:velo_prof}.

\begin{figure*}[t]
\begin{center}
	\includegraphics[width=17cm]{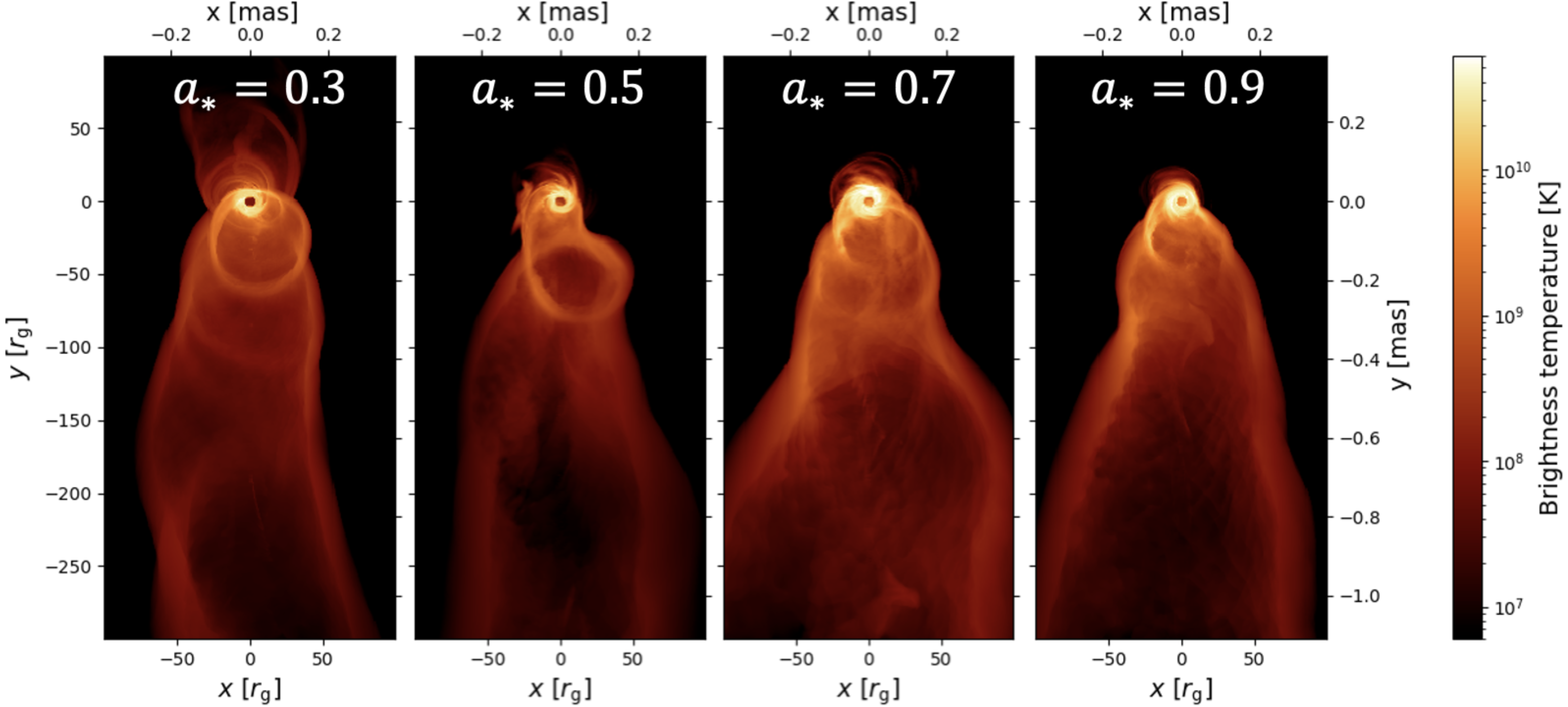}
\end{center}
    \caption{
    Slow-light snapshot images at 86 GHz for the four spin models with $a_* = 0.3, 0.5, 0.7,$ and $0.9$, from left to right. 
    The rightmost image for the highest spin ($a_* = 0.9$) is the same as the one shown on the left in Figure~\ref{fig:slow_fast}. 
    Movies are available in the online article and on \href{https://youtu.be/VIzLOhfJFaI}{YouTube}.
    }
    \label{fig:spins_slowlight}
\end{figure*}

Snapshot images for the four spin models, $a_* = 0.3$, 0.5, 0.7, and $0.9$, are shown in \figref{fig:spins_slowlight} (see also the online article or \href{https://youtu.be/VIzLOhfJFaI}{YouTube} for movies). 
All images exhibit a clear limb-brightening feature in the outer jet region around $y \lesssim -150~r_{\rm g}$, although the jet widths differ quantitatively. 
Meanwhile, the inner jet shows a qualitative difference: the three lower-spin models display distinct looped emission features, similar to those seen in the fast-light images but absent in the slow-light image for the highest spin. 
In addition, among the lower-spin cases, smaller spin values tend to produce the loop structures at larger (more downstream) regions along the jet.

The looped structures in the inner jet of the low-spin models can be interpreted in terms of plasma acceleration along the jet. 
As shown in \figref{fig:velo_prof}, slower spins systematically produce weaker acceleration, resulting in slower plasma bulk motion at all $z$. 
The jet in the highest-spin case ($a_* = 0.9$) undergoes the strongest acceleration and already becomes relativistic ($\gamma\beta = 1$, i.e., $\beta \approx 0.71$) at $z \sim 50~r_{\rm g}$, which corresponds to $y \approx -15~r_{\rm g}$ in the images. 
As described in Subsections~\ref{subsec:slow_fast} and~\ref{subsec:edges_loop}, the relativistic plasma bulk motion from this inner region all the way downstream produces a jet with two sharp edges across almost the entire image through the slow-light effect.

Meanwhile, the plasma bulk acceleration is relatively mild in the lower-spin cases. 
For example, in the lowest-spin model ($a_* = 0.3$), $\gamma\beta$ barely reaches unity even at $z = 1000~r_{\rm g}$. 
In such cases, the \textit{keeping up} of the plasma motion with the light rays is unlikely to occur; instead, the light rays overtake the plasma motion in the jet before reaching the screen. 
As a result, the jet exhibits looped structures at large distances downstream, reflecting the intrinsic features of the GRMHD model.

The moderate-spin model with $a_* = 0.5$ represents an intermediate regime in which the plasma velocity becomes relativistic ($\gamma\beta = 1$) at $z \sim 300~r_{\rm g}$. 
In this case, the transition from subrelativistic to relativistic regimes is directly imprinted on the jet image, appearing as a change from a loop-shaped structure in the inner region to a sharpened double-edged jet in the outer region. 
This transition occurs at around $y \sim -100~r_{\rm g}$, corresponding to the projected altitude of $z \sim 300~r_{\rm g}$ for an inclination of $i = 163^\circ$.

The second-highest spin model ($a_* = 0.7$) shows features quite similar to those of the highest-spin case, reaching the relativistic regime ($\gamma\beta = 1$) at a relatively small distance of $z \sim 100~r_{\rm g}$. 
In this model, a double-edged jet is persistently seen in the downstream region, while loop-like components occasionally appear closer to the BH at $y \gtrsim -50~r_{\rm g}$, which are scarcely seen in the highest-spin case.

In this way, the acceleration of the plasma up to the relativistic regime is manifested in slow-light images as a transition from loop-like structures in the inner jet {($\gamma\beta<1$)} to double-edged features in the outer jet {($\gamma\beta>1$)}. 
Since the plasma acceleration profile along the jet strongly and systematically depends on the BH spin (\autoref{fig:velo_prof}), observations of the image morphology in the jet-launching region can, in principle, provide constraints on the spin.

\subsection{Multi-Wavelength Jet Images}\label{subsec:mwl} 

\begin{figure*}
\begin{center}
    \includegraphics[width=18cm]{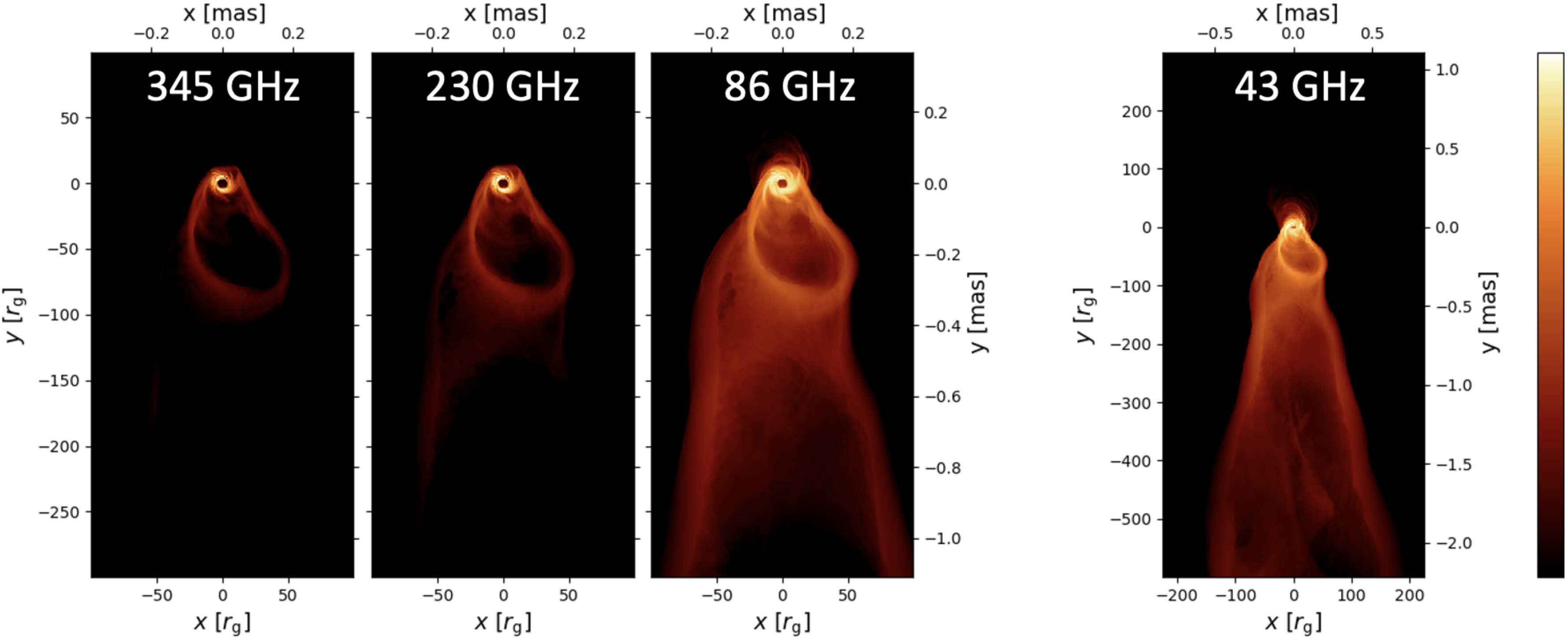}
    \includegraphics[width=18cm]{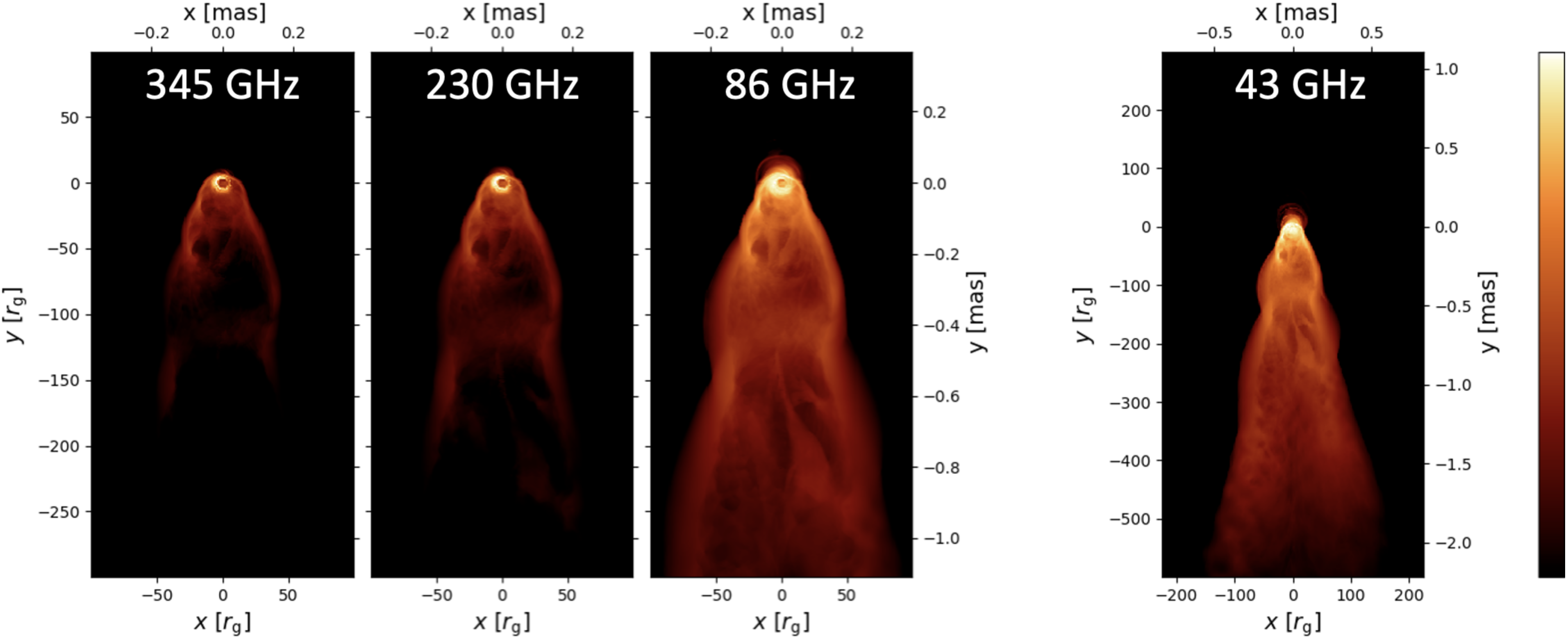}
\end{center}
    \caption{
    Snapshot images at 345, 230, 86, and 43~GHz, from left to right, for the $a_* = 0.5$ model (top row) and the $a_* = 0.9$ model (bottom row). 
    Note the difference in spatial scale between the left three images and the rightmost one. 
    Movies are available in the online article and on \href{https://youtu.be/VIzLOhfJFaI}{YouTube}.
    }
    \label{fig:a05_a09_mwl}
\end{figure*}

In the last subsection, we saw that the BH spin is imprinted on the jet image through the slow-light effect arising from the relativistic acceleration of plasma. 
Next, we discuss the possibility of constraining the spin using multi-scale, multi-wavelength images.

Snapshot images at 345, 230, 86, and 43~GHz are shown in Figure~\ref{fig:a05_a09_mwl} for the $a_* = 0.5$ and $0.9$ models. 
The 43~GHz images are calculated with a larger field of view extending to $y = -600~r_{\rm g} \approx 2.2~{\rm mas}$, compared to the other three frequencies, by placing the observer's screen at $r = 2600~r_{\rm g}$. 
Accordingly, the 43~GHz images are obtained at $t_{\rm scn,43} = t_{\rm scn} + 1300~t_{\rm g}$, accounting for the light-travel time delay between the screens at $r = 1300~r_{\rm g}$ (used for the other frequencies) and $r = 2600~r_{\rm g}$.\footnote{We note that this time delay simply arises from the difference in the setup of the observer’s screen and clock, and that all the images correspond to the same observational time on the Earth.}
Both models exhibit a double-edged jet in the outer region of $y < -100~r_{\rm g}$ down to $-600~\rg$, as clearly seen in the two low-frequency images. 
Meanwhile, as shown in the previous subsection, the low-spin model produces a loop-like feature in the inner region in the high-frequency images, in contrast to the persistent, sharp edges extending down to the ring emission around the BH in the high-spin case.

Based on these images, we can make predictions for future VLBI observations of the M87 jet. 
First, moving upstream along the jet as the angular resolution increases, the strongly accented, limb-brightened structure is expected to remain detectable up to milliarcsecond scales, which have already been observed at millimeter wavelengths \citep[e.g.,][]{2018A&A...616A.188K,2018ApJ...855..128W}. 
Next, two possible scenarios can be distinguished in the inner sub-mas region: one featuring loop-like structures (the $a_*=0.5$ model) and the other showing a persistent double-edged jet ($a_*=0.9$). 
This region is precisely the target of forthcoming global VLBI projects such as the EHT (\citealp{2024arXiv241002986T}), ngEHT (\citealp{2023Galax..11..107D}), and BHEX (\citealp{2024SPIE13092E..2DJ,2024SPIE13092E..2EA}). 
With their enhanced angular resolution and sensitivity, these instruments aim to detect the newly launched jet components from the innermost region near the BH and to image the jet and BH ring simultaneously. 
The resulting images might allow us to constrain the BH spin by identifying the location at which the loop-to-edge transition happens, as illustrated by the difference between the two rows in \figref{fig:a05_a09_mwl}.

\begin{figure*}
\begin{center}
	\includegraphics[width=15cm]{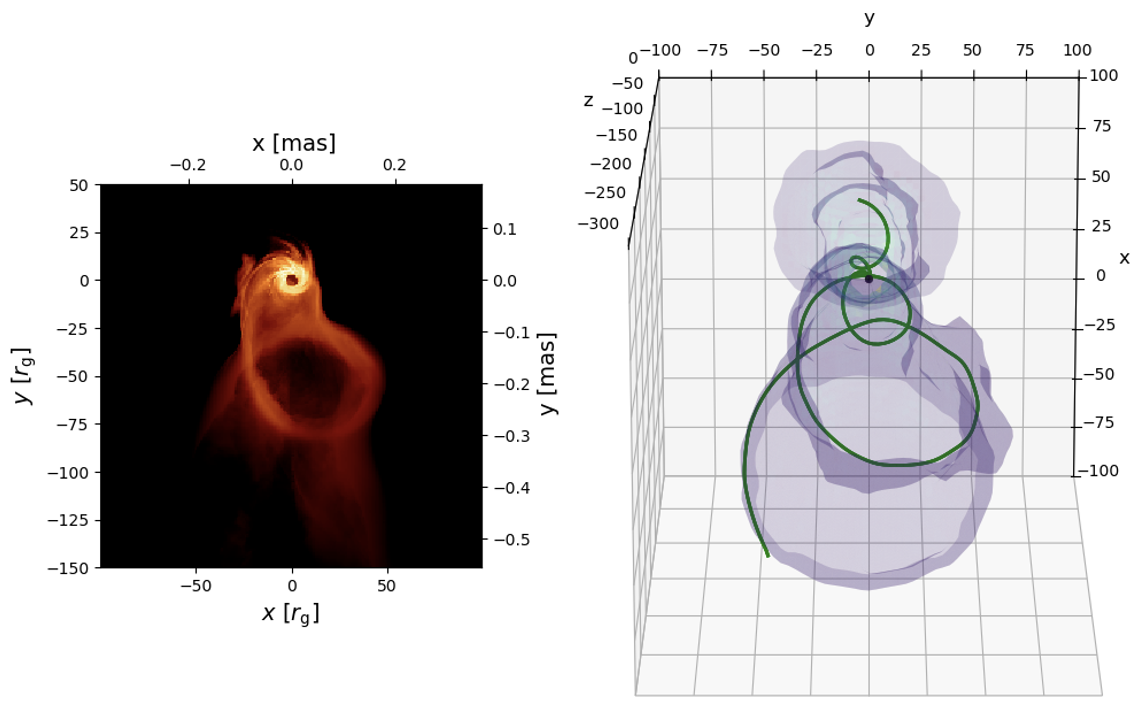}
\end{center}
    \caption{
    A snapshot image at 86 GHz for the $a_* = 0.5$ model (left) and a visualization of the $\sigma = 5$ surface and a magnetic field line in the corresponding GRMHD snapshot (right). 
    The magnetic field structure is taken from the GRMHD data at a representative epoch corresponding to the inner jet emission around $y \sim -50~r_{\rm g}$ in the image. 
    The field line originates at the outermost radius of the $\sigma = 5$ surface at $z = -300~r_{\rm g}$ and extends inward to $z = +100~r_{\rm g}$, remaining within the jet sheath. 
    The right rendering are viewed from an inclination angle of $i = 163^\circ$ with respect to the $z$-axis, aligned with the GRRT image on the left. 
    The left image is the same as the second panel from the left in Figure~\ref{fig:spins_slowlight}, but with a different color scale to emphasize the loop-like feature.
    }
    \label{fig:a05_Bline}
\end{figure*}

Here, we note that while we have so far selected representative images for each spin case, all models exhibit large variability reflecting that of the underlying MAD GRMHD simulations. 
As seen in the movies (see the online article or \href{https://youtu.be/VIzLOhfJFaI}{YouTube}), the width, extent, and position angle of the jet all vary on a timescale of $\sim 100~t_{\rm g}$, corresponding to roughly one month. 
In addition, the propagation of persistent jet components (e.g., loop features or the waist-like structures in the jet outline; see also Subsection~\ref{subsec:edges_loop}) can occasionally be seen. 
This time variability itself serves as an imprint of the spin, plasma velocity, and magnetic field configuration; for instance, slower jet propagation can be seen in the movies for the lower-spin models due to slower plasma bulk motion (see also Subsection~\ref{subsec:morphology}). 
In this regard, long-term monitoring of the M87 jet over several years with a monthly cadence will place strong constraints on the BH spin, and potentially provide important validation of the Blandford–Znajek process as the jet-driving mechanism.

\section{Discussion} \label{sec:discussion}

\subsection{Inner Loops and Magnetic Field Geometry}\label{subsec:Bfield}

In Subsections \ref{subsec:loop_edge} and \ref{subsec:mwl}, we showed that the velocity profile of the plasma motion in the jet is imprinted in the slow-light images as a transition from loop-like structures at slow speeds in the upstream region close to the BH to a double-edged feature at more relativistic speeds in the downstream jet. 
Here, we investigate the relationship between {these looped features} and the magnetic field structure in the jet.

As described in Subsection~\ref{subsec:edges_loop}, the double-edged appearance of the jet in the downstream region arises from the assumed anisotropic electron distribution \citep{2025ApJ...984...35T} coupled with the apparent longitudinal elongation and blurring of intrinsically looped jet components. 
It is therefore not easy to extract essential information about the jet structure from a slow-light image of the downstream jet. 
Here we focus on the loop-like emissions in the upstream region, which are less affected by plasma motion.

In Figure~\ref{fig:a05_Bline}, we present a visualization of GRMHD snapshot data for the $a_* = 0.5$ model. 
The visualization consists of a rendering of the magnetization contour surface at $\sigma = 5$ and a magnetic field line in the zero-angular-momentum observer (ZAMO) frame, aligned in time with the corresponding snapshot slow-light image from Figure~\ref{fig:spins_slowlight} (using a different color scale to highlight the loop-like features). 
The GRMHD snapshots are taken from a moment $t \approx t_{\rm scn} - 1130~t_{\rm g}$, when the light rays reaching $y \sim -50~r_{\rm g}$ on the image pass through the jet (see also \figref{fig:slow_fast}). 
The magnetic field line originates at the radius of the outermost point of the $\sigma = 5$ surface on the $(x,y)$ plane at $z = -300~r_{\rm g}$,\footnote{The loop-like features in the GRMHD models appear as bump-like structures on the iso-$\sigma$ surface.} 
and is traced inward until it reaches $z = +100~r_{\rm g}$, remaining within the jet-sheath region enclosed by the iso-$\sigma$ surface.

\begin{figure*}
\begin{center}
    \includegraphics[width=15cm]{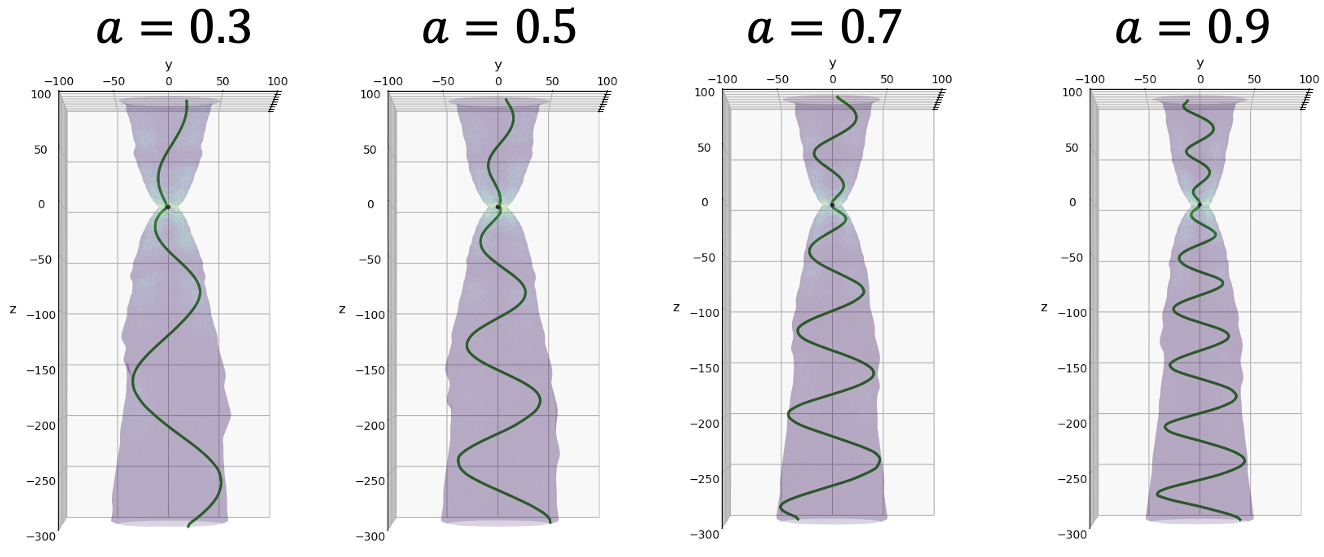}
\end{center}
    \caption{
    Same as the right panel of Figure~\ref{fig:a05_Bline}, but for a nearly edge-on view with an inclination angle of $i = 91^\circ$, using the GRMHD data averaged over $5000~t_{\rm g}$ for the four spin models.
    }
    \label{fig:spins_Bline}
\end{figure*}

It is clearly seen that the loop-like jet feature in the image follows a trajectory similar to that of the magnetic field line within the jet in the GRMHD model. 
In addition, the outline of the jet image traces the magnetic field line, exhibiting a characteristic ``waist'' at the crossing point of the magnetic loop. 
These results suggest that the loop-like structures in the GRMHD model and the corresponding emission features in the images reflect the geometry of a narrow bundle of magnetic field lines inside the jet, which is anchored close to the spinning BH and drives the plasma outward, presumably via the Blandford–Znajek process.
{Technically, the comparison shown in Figure~\ref{fig:a05_Bline} is not quite like-to-like, since the slow-light image in the left panel is based on data from multiple times whereas the GRMHD data in the right panel corresponds to a single instant of time. However, since we are focusing on the upstream region of the jet near the BH, where the plasma speed is fairly low (Figure~\ref{fig:velo_prof}), the patterns still match fairly well.} 

The correspondence between the loop-like features and the magnetic field lines provides valuable information about the BH spin. 
In Figure~\ref{fig:spins_Bline}, the $\sigma = 5$ surface and a representative magnetic field line are shown in the same manner as in Figure~\ref{fig:a05_Bline}, but for an inclination angle of $i = 91^\circ$, using GRMHD data averaged over $5000~t_{\rm g}$ for the four spin models. 
It is clearly seen that the number of magnetic field loops for a given distance along the jet increases with higher spin and is roughly proportional to the spin parameter $a_*$. 
This behavior is consistent with the predictions of force-free jet models, which give $\sharp(\mathrm{loops}) \propto B_\phi / (R B_z) \propto a_*$ (e.g., \citealp{2008MNRAS.388..551T}), where $B_\phi$ and $B_z$ are the toroidal and vertical components of the magnetic field, respectively, and $R$ is the cylindrical radius {(see \autoref{app:MagneticFieldWinding} for a brief derivation)}. 
This suggests that the looped magnetic field structures and the resulting inner loop-like emissions are formed by the frame-dragging effect, as described in the Blandford–Znajek process, and directly encode the BH spin.

Given the direct proportionality between the magnetic field winding rate and the BH spin, detection of such loop emission in images could prove valuable for constraining the spin and for testing the Blandford–Znajek process.  We note that low-spin cases are expected to be easier to measure in this regard, given that the transition to the double-edge morphology occurs at larger spatial scales for low-spin BHs.

\subsection{Jet Morphology and Variability}\label{subsec:morphology}

In this subsection, we examine some general features of jet images for different spin values by calculating time-averaged images over various time scales. 
Based on the results, we  perform in the next subsection a comparison with observations of the jet-launching region in M87.

\begin{figure*}
\begin{center}
    \includegraphics[width=18cm]{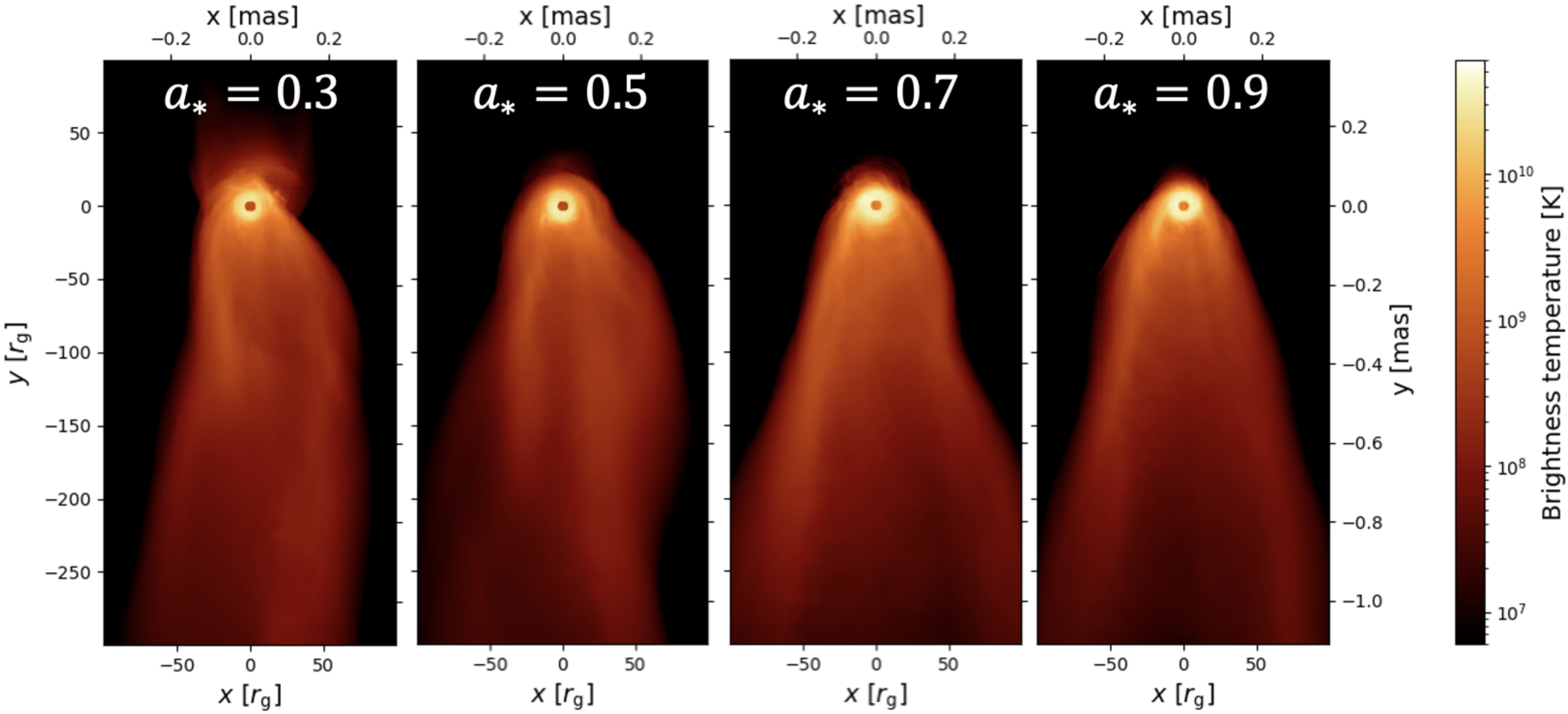}
    \includegraphics[width=18cm]{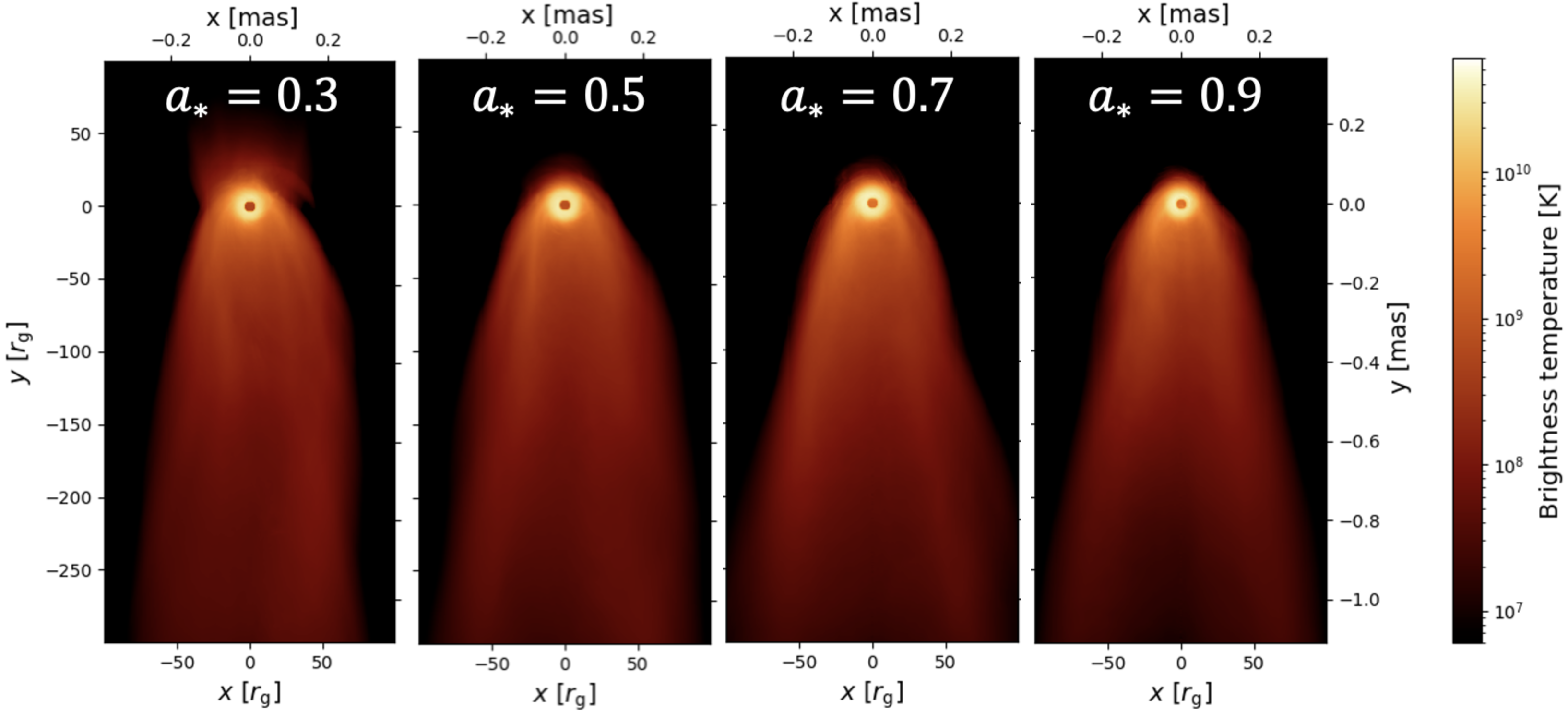}
\end{center}
    \caption{
    Time-averaged images at 86~GHz for the four spin models over two durations: $1000~t_{\rm g}$ (top) and $5000~t_{\rm g}$ (bottom), corresponding to approximately one and five years for M87, respectively.
    }
    \label{fig:1000average}
\end{figure*}

In the upper row of \figref{fig:1000average}, we show time-averaged images at 86~GHz for the four spin models, using an averaging time of $1000~t_{\rm g}$, which roughly corresponds to one year for M87. 
The time-averaged images can be divided into two groups based on their jet morphology: low-spin cases ($a_* = 0.3$ and $0.5$) and high-spin cases ($a_* = 0.7$ and $0.9$). 
Specifically, the two lower-spin models exhibit relatively narrow downstream jets with clear signs of wobbling or bending, even after averaging over a one-year timescale, whereas the higher-spin models produce wide and straight jets with no evidence of wobbling\footnote{{We note that the GRMHD simulations on which this work is based used ``reflecting boundary conditions" at the poles \citep{2022MNRAS.511.3795N}. These boundary conditions are standard in the field, but they might tend to damp down lateral motions of the jet and thereby artificially reduce the amplitude of any wobbles. It would be worthwhile to run the same models with ``transmitting boundary conditions" at the poles, which are expected to have a much weaker damping effect.}}.

The {wobbling} behavior of the jets with slowly spinning BHs can be explained by two factors. 
The first is that the jets intrinsically wobble on the scale of $z \lesssim 1000~r_{\rm g}$ in the GRMHD models, corresponding to $\lesssim 1~{\rm mas}$ in the images, as seen in the image movie (see the online article or \href{https://youtu.be/VIzLOhfJFaI}{YouTube}). 
The second factor is the relatively slow propagation of the jets, as shown in \figref{fig:velo_prof}. 
This naturally leads to slow convergence when time-averaging the images, combined with the weak slow-light effect that would otherwise elongate and smooth the jet appearance through plasma motion, as described in Subsection~\ref{subsec:slowlight}.\footnote{Indeed, we confirmed that, under the fast-light approximation, even the high-spin models show poor convergence in the $1000~t_{\rm g}$ averages, with non-systematic features remaining in the outer jet. 
This is because the plasma would propagate only $\sim 1000~r_{\rm g}$ even for $\beta = 1$, which corresponds to $\sim 290~r_{\rm g}$ in projection for $i = 163^\circ$, insufficient to reach the bottom edge of the image from the origin.}

The above difference between the slow- and fast-spin models disappears when the images are averaged over sufficiently long durations. 
In the bottom row of \figref{fig:1000average}, we present time-averaged images where the averaging time is now $5000~t_{\rm g}$, corresponding to roughly five years for M87. 
All the average images show good qualitative convergence toward a wide non-wobbly jet (compare also the two time-averaged images from the slow- and fast-light calculations in \figref{fig:average}). 
We note that the wide jets seen in the $5000~t_{\rm g}$-averaged images in the two low-spin models result from the wobbling motion of intrinsically narrow jets in the $1000~t_{\rm g}$ images being averaged out, whereas the two high-spin models stably produce wide and straight jets already with $1000~t_{\rm g}$ averaging, and even in the individual snapshot images shown in \autoref{fig:8snapshots}.

\subsection{Comparison with Large-Scale Observations}\label{subsec:obs_comp}

Finally, we compare the results in the last subsection with existing observations on large scales. 
\citet{2018A&A...616A.188K} presented a stacked 86~GHz image of the M87 jet, obtained from observations between 2004 and 2015, which can be directly compared with our $5000~t_{\rm g}$-averaged images. 
Their image shows a clear limb-brightening feature on sub-mas scales, without any loop or helical structure, whereas hints of such features occasionally appear in single-epoch images. 
This observed morphology is consistent with our four models, which exhibit inner loop-like features in snapshot images but double-edged jets in the time-averaged images.

Related to the wobbling jet, \citet{2023Natur.621..711C} detected a precession of the M87 jet with a period of about 11 years, based on observations spanning 22 years. 
In their analysis, one-year stacked images at 22–43~GHz on several mas scales show a clear bending of the jet as a manifestation of this precession. 
In this regard, our low-spin models may be favored over high-spin models, as they exhibit wobbling jet images in $1000~t_{\rm g}$ averages. 
However, as discussed by \citet{2023Natur.621..711C}, jet precession can also be triggered by the Lense–Thirring precession \citep{1918PhyZ...19..156L,1975ApJ...195L..65B} of a tilted accretion disk with respect to the BH spin axis, which would be more pronounced for high spin values \citep[see, e.g.,][]{2021MNRAS.507..983L}.

The observed wide jet in M87 weakly favors the high-spin models, which produce a wide jet on the downstream regions of $\sim 1-2~{\rm mas}$ over time, as also noted by \citet{2025ApJ...984...35T}. 
This may introduce a tension in the spin determination, given that the wobbling analysis in the previous paragraph favors lower spins.

A counter-jet has been detected in M87 \citep[e.g.,][]{2018ApJ...855..128W}.
Among our models, only the lowest-spin case ($a_* = 0.3$) produces bright emission from the counter jet in both the snapshot and time-averaged images. 
This originates from the slow acceleration in this low-spin model, where the plasma reaches only $\gamma\beta \sim 0.5$ around $z \sim 200~r_{\rm g}$, resulting in a significantly smaller intensity contrast between the approaching and receding jets than in the higher-spin models with twin relativistic ($\gamma\beta \gtrsim 1$) jets on this scale. 
More stringent constraints on the spin may be expected from future long-term observations with higher angular resolution and sensitivity.

To round out this discussion, we note that the jet in M87 is quite powerful, despite the low mass accretion rate of the BH. If the jet is produced by the Blandford-Znajek mechanism, models with rapidly spinning BHs may be preferred over slowly spinning BHs, owing to their higher jet power efficiency \citep{Tchekhovskoy2011,2022MNRAS.511.3795N}. 

\section{Conclusions}\label{sec:conclusions}

In this work, we have demonstrated that the slow-light approach in radiative transfer calculations -- which accounts for plasma evolution during the propagation of light rays -- is important for accurately modeling the jet-launching region near supermassive black holes, where the plasma fluid is accelerated and becomes relativistic.  We carry out a detailed investigation of the impact that slow-light has on simulated images of the jet-launching region in M87, for which the viewing angle is known to be small ($\sim$17\,degrees off from the line of sight) and thus plasma propagation along the same direction as emitted light rays is expected to result in prominent slow-light effects. We observe that the inclusion of slow-light leads to an apparent stretching and smoothing of intrinsically helical or loop-shaped jet components into a double-edged, cone-shaped emission structure, qualitatively consistent with the observed morphology of the M87 jet.

In more detail, we find that the use of slow-light results in a number of qualitative differences compared to fast-light images of the M87 jet:
\begin{itemize}
    \item \textbf{Superluminal motion}: Emission features in the slow-light movies appear to propagate downstream more rapidly than those in the fast-light movies, particularly at large radii where the plasma moves relativistically. 
    This arises because a single jet component can successively contribute to emission along multiple light rays as it propagates outward, producing radiation from progressively downstream regions at later times (Subsection \ref{subsec:slow_fast} and Figures~\ref{fig:slow_fast} and \ref{fig:pictures}). 
    As a result, the same physical structure in the jet appears to move faster across the image when viewed by a distant observer—a manifestation of the well-known superluminal motion effect.
    \item \textbf{Double-edged structure}: Slow-light jet images exhibit more limb-dominated structures than their fast-light counterparts, with the degree of limb dominance increasing toward larger radii. 
    This behavior arises because a single jet component can contribute to emission along multiple light rays as it propagates downstream, as described above, while simultaneously nearly “keeping up” with individual light rays and continuously contributing to the radiative transfer process along them. 
    These combined effects stretch and smooth the intrinsically loop-shaped emission structures into an apparently hollow conical geometry, which, in combination with anisotropic emission concentrated along the magnetic field lines, increases the optical depth along the light rays passing through the two edges (Subsection \ref{subsec:edges_loop} and Figures~\ref{fig:8snapshots} and \ref{fig:a03_slow_fast}).
    \item \textbf{Stability of jet structure}: Slow-light movies show less propensity for “wobbling” of the jet compared to fast-light movies. 
    This behavior can be understood as another consequence of the stretching and smoothing effect on the intrinsically loop-shaped emission structures, which tend to induce more variability in the transverse direction in the fast-light images (Figure~\ref{fig:8snapshots}; see also movies on the online article or \href{https://youtu.be/VIzLOhfJFaI}{YouTube}).

\end{itemize}
\noindent We confirm prior results that regions of the image where the plasma is primarily moving sub-relativistically (e.g., the innermost jet) are not heavily impacted by the use of slow-light. 
We also demonstrate that both slow-light and fast-light images for generally optically thin cases become nearly identical when averaged over sufficiently long timescales (Subsection \ref{subsec:average}, Figure~\ref{fig:average}).

\bigskip

{We} surveyed the slow-light effect across a range of BH spin values and investigated the relationships among the resulting images, plasma acceleration, and magnetic field structure. 
Focusing on the potential to determine the BH spin and to test the Blandford–Znajek process through future observations, we summarize our main findings as follows.

\begin{itemize}
    \item {\bf Loop–edge transition:} The transition of the observed image from a loop-like to double-edged structure occurs where the plasma bulk motion becomes relativistic, roughly where $\gamma\beta\sim1$. 
    Since the plasma acceleration profile systematically depends on the BH spin (the higher the spin, the stronger the acceleration; see \figref{fig:velo_prof}), 
    the location of this loop–edge transition can serve as a powerful diagnostic of the BH spin from jet images (Subsection~\ref{subsec:loop_edge}, Figure~\ref{fig:spins_slowlight}). 
    We found that, when observing from downstream to upstream with increasing angular resolution (by going to increasingly higher obsserving frequencies), the jet persistently exhibits a double-edged structure up to $\sim$1~mas. 
    However, entering the sub-mas scale, one may detect loop-like emission in low-spin cases or a persistent double-edged morphology in high-spin cases (Subsection~\ref{subsec:mwl}, Figure~\ref{fig:a05_a09_mwl}).

    \item {\bf Inner loops and magnetic field geometry:} The loop-shaped emission structures in the inner jet directly trace the trajectories of magnetic field lines within the jet (Figure~\ref{fig:a05_Bline}), which also show systematic dependence on the spin (the higher the spin, the {faster the winding rate of the loops}; see \figref{fig:spins_Bline}). 
    This correspondence is particularly clear in the lower-spin cases, where the inner loops are visible {farther down the jet}, 
    while higher angular resolution and sensitivity are required to capture the loops in the innermost region for higher spins (\autoref{subsec:Bfield}). 

    \item {\bf Jet morphology and variability:} The low-spin models exhibit jet wobbling in the movie sequences and show a bending jet structure even after averaging over $1000~t_{\rm g}$ (roughly one year), consistent with the mas-scale bending observed in M87. 
    In contrast, the high-spin models persistently produce a straight and wide jet, consistent with observations on larger scales. 
    We also found that only the lowest-spin case exhibits a bright counter jet, owing to {weaker} plasma acceleration and consequently {lower} contrast with the approaching jet due to relativistic beaming (\autoref{subsec:obs_comp}). 
    These model features create some tension in spin determination, underscoring the need for comparison with high-resolution and high-sensitivity observations.
\end{itemize}

As noted in the last point, a wide range of BH spin values remain viable -- or in tension -- when compared with existing large-scale observations of the M87 jet. 
The movies also show that all image features, such as the trajectories of loops, the width of the double edges, and the extent and position angle of the overall jet, exhibit significant variability on week-long timescales, reflecting the dynamics of the underlying MAD models. 
Long-duration monitoring observations by future instruments will therefore be essential for placing strong constraints on the BH spin and testing the magnetically driven mechanism of jet formation.

\bigskip

\begin{acknowledgments}
The authors thank Zachary Gelles, Andrew Chael, Charles Gammie, and Eliot Quataert for constructive discussions. 
YT is grateful for support from JSPS (Japan Society for the Promotion of Science) Overseas Research Fellowship, the National Natural Science Foundation of China (Grant No. 12325302, 11933007), and the Shanghai Pilot Program for Basic Research, Chinese Academy of Sciences, Shanghai Branch (JCYJ-SHFY-2021-013). 
{Support for DWP was provided by the NSF through grants AST-1935980, AST-2034306, and AST-2535855; and by the Gordon and Betty Moore Foundation through grants GBMF5278 and GBMF10423.}
This work was supported in part by the Black Hole Initiative at Harvard University, which is funded by the John Templeton Foundation (grants 60477, 61479, and 62286) and the Gordon and Betty Moore Foundation (grant GBMF8273). 

\end{acknowledgments}
%
\vspace{5mm}





\bigskip

\appendix

\section{Survey of GRMHD Snapshot Cadence}\label{apdx:cad}

\begin{figure*}
\begin{center}
	\includegraphics[width=18cm]{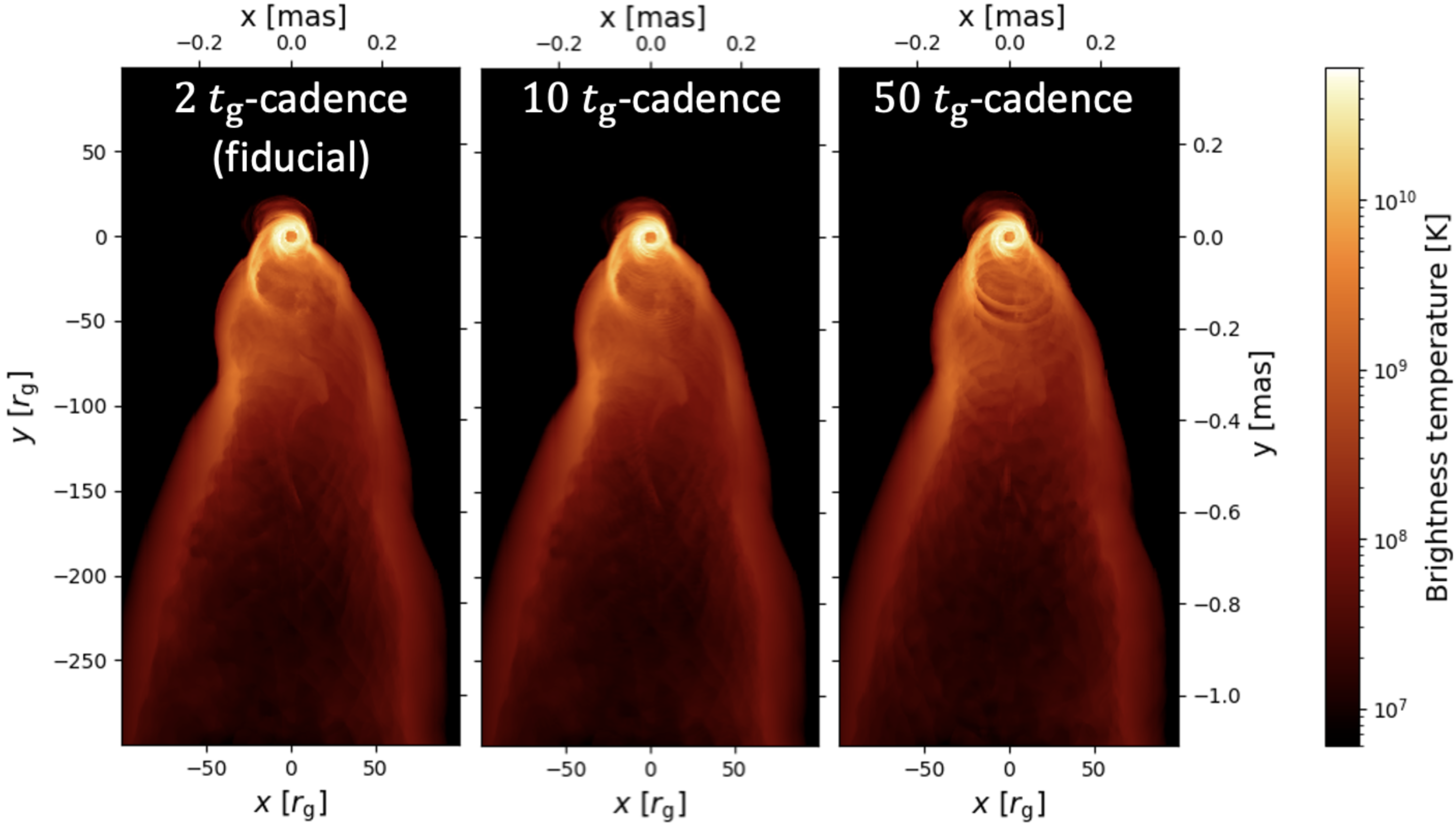}
\end{center}
    \caption{
    Slow-light snapshot images at 86 GHz for the $a_* = 0.9$ model with different GRMHD snapshot cadences of $2~\tg$, $10~\tg$, and $50~\tg$, left to right.
    }
    \label{fig:cadence}
\end{figure*}

Here, we examine the cadence of GRMHD snapshot data used in the slow-light calculations, which was set to $2~t_{\rm g}$ in the main text. 
In \figref{fig:cadence}, three slow-light images are shown for cadences of $t_{\rm cad} = 2~t_{\rm g}$ (same as in the right panel of \figref{fig:slow_fast}), $10~t_{\rm g}$, and $50~t_{\rm g}$. 

We find that the larger two cadences introduce artificial discontinuities in the emission from a persistent jet component, on a spatial scale of roughly $\sim \beta c \Delta t_{\rm cad} \sin(163^\circ)$. 
For example, in the case with the largest cadence of $\Delta t_{\rm cad} = 50~t_{\rm g}$, the loop- or arch-shaped feature at $y = -50~r_{\rm g}$ in the other two images is split into two components separated by about $10~r_{\rm g}$ on the image, originating from two temporally adjacent GRMHD snapshots at this large cadence. 
Such artificial discontinuities appear broadly in the inner region ($y > -150~r_{\rm g}$).

The image with a moderate cadence of $\Delta t_{\rm cad} = 10~t_{\rm g}$ shows a structure quite similar to that obtained with the smaller cadence. 
However, a careful inspection reveals a slight discontinuity on a tiny scale of $\sim 1~r_{\rm g}$ in the inner region ($y > -50~r_{\rm g}$). 
Therefore, we may conclude that $\Delta t_{\rm cad} = 10~t_{\rm g}$ is sufficient for analyses of the outer jet, whereas a cadence of $\Delta t_{\rm cad} = 2~t_{\rm g}$ or shorter is required when focusing on the fine features around the photon ring and the BH shadow.

\section{Isotropic Synchrotron-Electrons and Slow Light}\label{apdx:iso}

\begin{figure*}
\begin{center}
    \includegraphics[width=10cm]{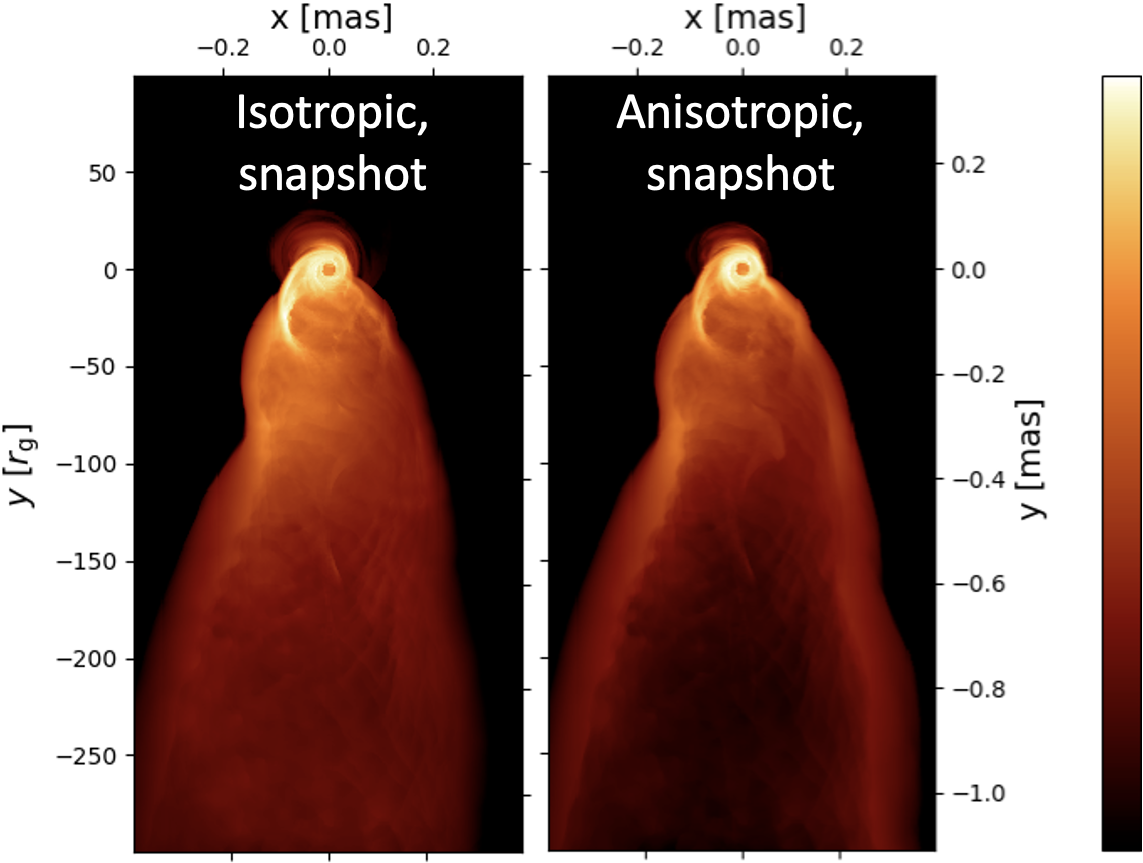}
	\includegraphics[width=10cm]{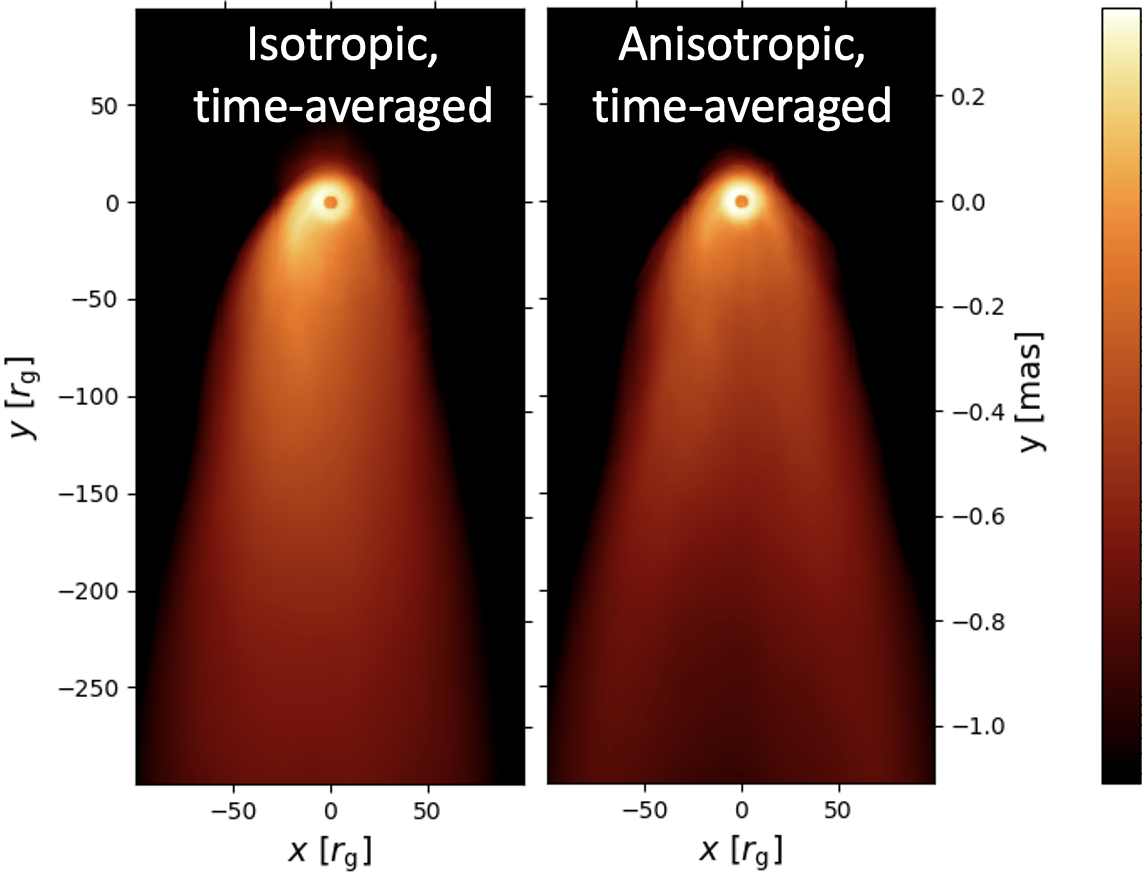}
\end{center}
    \caption{
    Top: Slow-light snapshot images at 86 GHz for the $a_* = 0.9$ model with isotropic (left) and anisotropic (right) distributions of nonthermal electrons in the jet region. 
    The anisotropic image is identical to those shown in Figures~\ref{fig:slow_fast} and~\ref{fig:spins_slowlight}. 
    Bottom: Same as the top panels, but showing time-averaged images over a duration of $5000~t_{\rm g}$. 
    The anisotropic image is identical to that in Figure~\ref{fig:average}.
    }
    \label{fig:iso}
\end{figure*}

In Figure~\ref{fig:iso}, snapshot and time-averaged images for the $a_* = 0.9$ model with isotropic nonthermal electrons ($\eta = 1$ in Equation~\ref{eq:phi}; see \citealp{2025ApJ...984...35T} for details) are shown together with those for the anisotropic case. 
The isotropic case does not exhibit the double-edged, limb-brightened feature but instead shows a single-edged or ridge-brightened structure, due to strong emission in directions perpendicular to the magnetic fields and to the relativistic Doppler beaming effect.
Meanwhile, the effect of slow light can also be seen in the isotropic case. 
In the downstream region ($y < -50~r_{\rm g}$), the jet remains persistently brighter on the right side without showing any trace of loops, owing to the stretching and smoothing caused by the nearly parallel motion of the plasma bulk and light rays, as also seen in the anisotropic case.

Based on the above results, the limb-brightening in the jet-launching region can be summarized as follows. 
First, the anisotropy in the pitch-angle distribution of electrons in the jet produces stronger emission along the two edges than in the interior ridge region. 
In the inner jet region, where the plasma bulk motion is subrelativistic, we can observe intrinsically loop-shaped emission that reflects the magnetic field geometry and is enhanced along the edges of the jet. 
However, in the outer, downstream region where the plasma motion becomes relativistic, a double-edged, prominently limb-brightened jet appears as a result of the combined effects of anisotropic emission and the stretching and smoothing caused by the slow-light effect.

\section{Magnetic field winding in force-free models} \label{app:MagneticFieldWinding}

Models of the jet structure that use force-free electrodynamics \citep[e.g.,][]{2008MNRAS.388..551T,2009ApJ...697.1164B,2025ApJ...981..204G} can provide a useful analytic framework for understanding the features seen in GRMHD simulations.  In this section, we provide a short derivation of the magnetic field loop winding rate described in \autoref{subsec:Bfield}.

At each point along any magnetic field line, the local line element $d\boldsymbol{\ell}$ must be parallel to $\boldsymbol{B}$, which means that we have the relationship

\begin{equation}
\frac{dr}{B_r} = \frac{r \sin(\theta) d\phi}{B_{\phi}} . \label{eqn:LineElement}
\end{equation}

\noindent To understand the oscillatory behavior of the field lines seen in \autoref{fig:spins_Bline}, we want to look at $\phi(r)$.  We start with

\begin{equation}
\frac{d\phi}{dr} = \frac{B_{\phi}}{r \sin(\theta) B_r} , \label{eqn:dphi_dr_1}
\end{equation}

\noindent which is just a rearrangement of \autoref{eqn:LineElement}. Expressions for $B_{\phi}$ and $B_r$ have been determined by \cite{2008MNRAS.388..551T}, who provides

\begin{align}
B_r & = r^{\nu - 2} \\
B_{\phi} & = - 2 \Omega r^{\nu - 1} \tan\left( \frac{\theta}{2} \right) .
\end{align}

\noindent where $\nu$ controls the collimation rate of the jet and $\Omega$ is the field line rotation rate (set primarily by the spin of the black hole).  Plugging these expressions into \autoref{eqn:dphi_dr_1} yields

\begin{equation}
\frac{d\phi}{dr} = - \Omega \sec^2\left( \frac{\theta}{2} \right) .
\end{equation}

\noindent We can use the fact that the magnetic stream function

\begin{equation}
\psi(r,\theta) = r^{\nu} \Big( 1 - \cos(\theta) \Big)  \qquad \Longrightarrow \qquad \psi_0 \equiv r_0^{\nu} \Big( 1 - \cos(\theta_0) \Big)
\end{equation}

\noindent is constant along a field line to replace $\theta$ with $r$, yielding

\begin{equation}
\frac{d\phi}{dr} = - \frac{\Omega}{1 - \frac{\psi_0}{2 r^{\nu}}} .
\end{equation}

\noindent For arbitrary values of $\nu$, the solution to the above equation can be expressed in terms of hypergeometric functions.  We can approximate the radial separation $\Delta r$ between consecutive magnetic field windings using

\begin{equation}
\Delta r \approx \frac{2 \pi}{|d\phi/dr|} = \frac{2 \pi}{\Omega} \left( 1 - \frac{\psi_0}{2 r^{\nu}} \right) .
\label{eq:Deltar}
\end{equation}

\noindent So the spatial period $\Delta r$ is inversely proportional to $\Omega$, approaching $2 \pi / \Omega$ as $r \rightarrow \infty$.  The number of loops seen in any particular interval along the jet will then be proportional to $r/\Delta r \propto \Omega \propto a_*$, as described in \autoref{subsec:Bfield}. 

To be more quantitative,  magnetic field lines in the jet, which are anchored on the BH horizon, have an angular velocity $\Omega$ roughly equal to 30-50\% of the angular velocity $\Omega_H$ of the BH horizon \citep[e.g.,][]{McKinney2007}. Noting that (in units of $1/t_g$)
\begin{equation}
    \Omega_H = \frac{a_*}{2(1+\sqrt{1-a_*^2})} \approx \frac{a_*}{4},
\end{equation}
and ignoring the factor in parentheses in Equation~(\ref{eq:Deltar}), we estimate the number of loops out to a distance $z=300r_g$ to be
\begin{equation}
    n_{\rm loops} = \frac{300r_g}{\Delta r} \approx 5a_*,
\end{equation}
which is roughly consistent with
Figure~\ref{fig:spins_Bline}.


\bigskip

\bibliography{slowlight}{}
\bibliographystyle{aasjournalv7}



\end{document}